\documentclass[symmetry,article,accept,pdftex,moreauthors]{mdpi} 
\usepackage{comment}

\usepackage{graphicx}
\usepackage{dcolumn}
\usepackage{bm}
\usepackage{xcolor}
\usepackage{soul}
\usepackage{pifont}

\usepackage{booktabs} 
\usepackage{multirow}

\newcommand{\cmark}{\ding{51}}%
\newcommand{\xmark}{\ding{55}}%

\firstpage{1} 
\pubvolume{18}
\issuenum{1}
\articlenumber{27}
\pubyear{2025}
\copyrightyear{2025}
\externaleditor{Stefano Profumo and Charalampos Moustakidis} 
\datereceived{21 October 2025} 
\daterevised{27 November 2025} 
\dateaccepted{12 December 2025} 
\datepublished{23 December 2025} 
\pdfoutput=1 

\Title{How Neutron Star Observations Point Towards Exotic Matter: Existing Explanations and a Prospective Proposal} 

\Author{Mauro Mariani 
 $^{1,2,}$*\orcidA{} and Ignacio F. Ranea-Sandoval 
 $^{1,2}$\orcidB{}}

\AuthorNames{Mauro Mariani and Ignacio F. Ranea-Sandoval}

\address{%
$^{1}$ \quad Grupo de Astrofísica de Remanentes Compactos, Facultad de Ciencias Astronómicas y Geofísicas, Universidad Nacional de La Plata, Paseo del Bosque S/N, La Plata 1900, Argentina;  iranea@fcaglp.unlp.edu.ar
\\
$^{2}$ \quad {Consejo Nacional de Investigaciones Cient\'ificas y T\'ecnicas} 
, Godoy Cruz 2290, \linebreak  {Ciudad Aut\'onoma de Buenos Aires 1425}, 
 Argentina}

\corres{Correspondence: mmariani@fcaglp.unlp.edu.ar}

\abstract{Multi-messenger astronomical observations of neutron stars, together with more precise calculations and constraints coming from dense matter microphysics, are generating tension with regard to equations of state models used to describe neutron star cores. Assuming an abrupt first-order phase transition with a slow conversion speed between phases, we propose different slow stable hybrid star configurations aiming to reconcile all current constraints simultaneously; within this framework, we also introduce a novel non-CSS parametrization to the quark matter equation of state and discuss its strengths and limitations. We analyze our model results in conjunction with a review of other relevant theoretical possibilities existing in the literature. We found that modern neutron star observations seem to favor the existence of some type of exotic matter in the neutron star cores; in particular, our slow stable hybrid star scenario remains a proposal capable of satisfying these constraints. However, due both to the existing skepticism regarding some of the adopted hypotheses in most extreme neutron star measurements and to the precise adjustment needed for the equation-of-state parameters, significant tension and open questions~remain.}


\begin{document}


\section{Introduction} \label{sec:intro}

Strong observational evidence of neutron stars (NSs) is available for almost sixty years since the breakthrough observation made by Jocelyn Bell in 1967~\cite{Hewish:1968ofa}. NSs, ultra-dense compact objects with enormous magnetic fields and very fast rotating speeds, are associated with extremely energetic events, such as radio pulsars, soft gamma repeaters, and anomalous X-ray pulsars (see, for example, Refs.~\cite{ozel:2016mra,vidana:2018asw,kondratyev:2019mom,esposito:2020mas} and references therein). In the last 15 years, several estimations of the masses and radii of NSs have become available. Observations of NSs are being made through pulsar timing in binary systems, such as the $2 M_\odot$ pulsars~\mbox{\citep{Demorest:2010sdm,Antoniadis:2013amp, Arzoumanian:2018tny, Cromartie:2020rsd, Fonseca:2021rfa}}; NICER observations of isolated NS~\citep{Miller:2019pjm,Riley2019anv,Miller:2021tro,Riley:2021anv, Salmi:2024anv, Choudhury2024anv,Mauviard:2025anv}; and multimessenger astronomy after GW170817~\citep{Abbott:2017oog, Abbott:2018gmo,Abbott:2020goo}. There also exist spectroscopic and photometric observational techniques for pulsars in binary systems, such as the \textit{black widow
} PSR~J0952-0607, with a very high mass estimation, $M = 2.35 \pm 0.17 M_\odot$~\citep{Romani:2022pjt}. In addition, two recent estimations of the mass and radius of compact objects have appeared as follows: the very low mass object HESS~J1731-347 ($M= 0.77^{+0.20}_{-0.17}\,M_\odot$ and $R= 10.4^{+0.86}_{-0.78}$ km)~\citep{Doroshenko:2022asl} and XTE~J1814-338, the mass of which is not so low ($M = 1.21\pm0.05\, M_\odot$) but it has a very small radius (\mbox{$R=7.0\pm0.4$ km})~\citep{Kini:2024ctp}. Besides these astronomical observations, other constraints on the equation of state (EOS) of dense matter arise from chiral effective field theory ({$\chi$EFT}) calculations (see, for example,~\citet{Drischler:2021lma} and references therein) and perturbative {Quantum Chromodynamics} (pQCD) (see, for example,  Ref.~\citep{Annala:2020efq} and references therein). All these estimations, when considered simultaneously, produce tension with the theoretical models for cold dense matter used to describe matter in the inner core of NSs.

Regarding the uncertainties over recent observations, we can particularly mention that changes to the atmospheric model used to determine the restrictions for HESS~J1731-347 produce a higher value for the mass, thus releasing some of the tension with such a low value. This issue has already been stated in Ref.~\cite{Doroshenko:2022asl} and also discussed in Ref.~\cite{alford:2023dcc}. Something similar happens for object XTE~J1814-338, where our poor understanding of the bursting mechanism might be introducing systematic errors that lead to such an \textit{extreme} object. In this case, though not preferred, an alternative model---also presented in Ref.~\cite{Kini:2024ctp}---produces different estimations of the mass and radius of this compact star {($M \sim 1.4\,M_\odot$ and \mbox{$R \sim 10$ km}). In this scenario, less tension with other restrictions is produced}. In relation to the challenging estimations of the parameters of PSR~J1231-1411, the lack of a reliable estimation for the distance and less restrictive priors on mass and observer inclination are certainly plotting against the determination of its mass and radius. The PSR~J0952-0607 mass measurement also introduces a very high uncertainty with respect to other binary massive NSs mass detections~\citep{Malik:2025oco}. As we will discuss, the mentioned tension is not necessarily relieved if these constraints upon which a cloak of distrust exists are rectified.

It is important to recall that there are no experimental facilities on Earth capable of studying matter in the density range between $\chi$EFT and pQCD calculations for cold dense matter. Moreover, theoretical schemes to construct its EOS from first principles systematically fail in this regime. For these reasons, there are large uncertainties regarding the behavior of matter in this range of densities. In this context, the most promising possibilities for understanding matter at these extreme densities are related to astronomical observations of NSs. In this sense, an important approach to addressing this issue is to produce model-independent studies that allow us to obtain a broad set of results to test against modern astronomical data. Such results are also important in the context of future astronomical observations that will become available with, for example, third-generation gravitational-wave detectors---such as the Einstein Telescope or Cosmic Explorer (see, for example, Refs.~\cite{punturo:2010tet,reitze:2019cet,Luck:2020tgg}).

Different models have been proposed to explain the complex scenario of astronomical observations. The measurements of the $2\,M_\odot$ pulsars created great tension with hadronic models as the inclusion---inevitable as energy density increases---of heavier baryons in the EOS {conflicts} 
 with such massive compact stars. This problem is commonly referred to as the \textit{hyperon puzzle} (see, for example, Refs.~\cite{bombaci:2017thp,ye:2025hds} and references therein). One of the most explored alternatives is the inclusion of deconfined quark matter inside these compact objects. We must stress that quark degrees of freedom are allowed by QCD, where the theoretical QCD phase diagram predicts the occurrence of a hadron-quark (HQ) phase transition for an unknown density at $T=0$; there also exist results that suggest that the existence of this kind of matter inside NSs is favored~\citep{Annala:2020efq}. For this reason, it is fair to include deconfined quark matter in the unknown EOS that describes matter in the inner depths of compact objects. EOSs allowing a phase transition involving quark deconfinement are commonly known as hybrid EOSs, which produce hybrid stars (HSs). 

The nature of this hypothetical HQ phase transition at low temperature is a matter of debate. For a long time, and continuing up to the present, many authors have studied the possibility of a first-order phase transition, both in QCD phase diagram studies~\citep{Halasz:1998pfd, Stephanov:2006qpd, Guenther:2020oot, Chen:2024pot, Lu:2025rtf} and NSs studies~\cite{Glendenning:1992nsc, Heiselberg:1999pti, Reddy:2000fop, Chamel:2013pti, Komoltsev:2024fop}; in the few last years, other authors argued that it could be a smooth \textit{crossover}, analogous to what occurs in the low-density and hight-temperature regime of the QCD phase diagram ~\cite{Schafer:1999coq, Masuda:2013hqc, Hirono:2019qhc, Brandes:2023eaa, Fujimoto:2025shq, Fukushima2025qpd}. Furthermore, other kinds of scenarios have also been proposed, such as the occurrence of intermediate phases with partial quark deconfinement, given by quarkyonic or \textit{spaghetti of quarks with glueballs} phases~\cite{Fujimoto:2025nso}. 
 Far from being settled, the question concerning the nature of the HQ phase transition at low temperature remains open and subject to diverse viewpoints; studying NSs as astrophysical laboratories could help to clarify this issue as more advanced and precise observations become available.

Moreover, within the first-order scenario, two limiting cases to describe such phase transitions are available in the literature as follows: hybrid EOSs in which a sharp transition, characterized by a gap in the energy density, $\Delta \varepsilon$, occurs at a given transition pressure, $P_t$, or phase transitions in which a mixed phase, where hadrons and quarks coexist, is formed. In the first case, the sharp phase transition is described using the Maxwell formalism (see, for example, Refs.~\cite{Baym:2018fht,Orsaria:2019pti} for the \textit{canonical} treatment in which solutions to different Lagrangians are stitched together and Refs.~\cite{Dexheimer:2010ana,Kumar:2024mna,celi:2025etr} for unified models). In the second case, the soft phase transitions with a mixed phase are constructed using the (bulk) Gibbs formalism (see, for example, Refs.~\cite{Logoteta:2022ieo,Constantinou:2023ffp} and references therein). Moreover, in this last scenario, geometrical structures can appear as a consequence of the interplay of the Coulomb and nuclear surface energies (see, for example, Refs.~\cite{Ju:2021hqp,Mariani:2024qhp} and references therein). The main physical quantity that determines the nature of the HQ phase transition is the (highly) unknown value of the HQ surface tension (see, for example, Refs.~\cite{pinto:2012stq,mintz:2013pda} and references therein). Moreover, if the sharp HQ phase transition is favored, the following additional key aspect needs to be taken into account: the conversion timescale between hadrons and quarks. This feature is of primary relevance, as it has been shown that if this timescale is \textit{slow} compared to the typical values of the fundamental mode of radial perturbations, a new branch of \textit{slow} stable HSs (SSHS) can exist where the condition $\partial M/\partial \varepsilon _c > 0$ is not satisfied, as proven in Ref.~\cite{Pereira:2017pte} (It is important to stress that stability in these extended branches also appears in different astronomical scenarios: electrically charged 
quark stars~\citep{Arbanil:2015eas}; anisotropic compact stars~\citep{mohanty:2024uan}; multiple-fluid compact objects~\citep{kain:2020roa,caballero:2024rms}; HSs with elastic phases in their cores~\citep{pereira:2021peq}; and hadronic NSs when perturbations are not assumed to preserve chemical equilibrium but \textit{frozen} populations of particles are considered for the perturbation~\citep{canullan:2025nss}.). On the contrary, if the timescale is \textit{rapid}, the traditional stability result remains valid, where the stability turning points lie at the critical points $\partial M/\partial \varepsilon _c = 0$. 

{Several theoretical frameworks---including, but not restricted to, HQ HSs---have been proposed in the literature to account for the observational and theoretical constraints related to compact objects. We present them briefly in order to discuss them in detail in the following sections. The possibility of two coexisting families of compact objects with different natures has been proposed in the works by~\citet{Drago:2014cvc}, \linebreak \citet{DiClemente:2024itc,Drago:2016tso1,Drago:2016tso2}. This \textit{two-family scenario} explains the high-mass objects as strange quark stars (QSs) and the low-mass and small-radii objects as hadronic NSs. Totally stable HSs (i.e., HSs that are stable in both \textit{slow}} and \textit{rapid} conversion regimes) \textls[+15]{with sharp interfaces and HSs where a mixed phase is formed are used in the works by~\citet{abgaryan2018two,ranea:2019eoh,carlomagno:2024tts,alvarez:2019tfo,Mariani:2024qhp}, \mbox{\citet{Pal:2025cah}}.} Within the same theoretical approach, there are also works presenting the impact of the occurrence of sequential phase transitions (a HQ one and a second one between different quark phases) in the inner core of compact stars, as presented in the works by~\mbox{\citet{alford:2017csw}},\citet{rodriguez:2021hsw,li:2023rhs}. Within the slow conversion regime, SSHSs are good candidates to help explain every constraint (see, for example, the works of~\citet{Goncalves:2022ios, Lugones:2023ama,Rau:2023neo}, \mbox{\citet{Rau:2023tfo}}, \citet{Mariani:2024cas,Laskos:2025xja}). Within this theoretical approach, slow QSs including phase transitions between different quark phases have also been recently proposed by~\citet{Zhang:2025sss, Yang:2025hqs}. In the work by~\citet{Sagun:2023wit}, a summary of different proposals, which include a dark matter (DM) contribution to NSs and HSs, is~presented.

Given the current scenario of recent observations and constraints, and the different theoretical proposals, the general objective of this work is to present an updated SSHS approach to explain the current tension among modern constraints and to analyze it in the context of the other proposals existing in the literature. Given the aim of exploring the SSHS scenario, we will construct HS configurations within the first-order slow phase transition hypothesis. In this context, the first part of this work presents novel results obtained from a model-independent perspective, including a new phenomenological non-constant speed of sound (CSS) parametrization that may be helpful for constructing hybrid EOSs consistent with microscopic results from pQCD. The second part, motivated by the aforementioned tension, reviews several theoretical approaches proposed to account for the full set of observations, discussing their strengths and limitations and comparing them with our own~proposal.

The present work is organized in the following manner. In Section~\ref{sec:eos}, we present the hybrid EOSs used to describe dense matter, including a detailed presentation of the novel non-CSS parametrization for quark matter. The results obtained from implementing these EOSs are presented in Section~\ref{sec:results}, where we also provide an astrophysical interpretation and compare them with other proposals in the literature. Section~\ref{sec:conclusion} is devoted to summarizing our main findings and presenting the conclusions. In Appendix~\ref{app:A} we present tables with detailed quantities of relevant stellar configurations obtained using with the different EOSs used in this work. We use the geometric unit system, where the gravitational constant and the speed of light are $G=c=1$.

\section{Hybrid EOS and HS Models} \label{sec:eos}

In order to obtain results with a model-independent approach, we adopt generic parametric models to construct both the hadron and quark phases of the proposed hybrid EOSs. In this sense, our approach is a continuation of the previous studies presented in Refs.~\citep{Lugones:2023ama, Mariani:2024cas}. However, in this work, we do not model, as in the previous ones, only HQ HSs, we also consider the possibility of quark--quark (QQ) HSs, in which a phase transition between two different quark phases is proposed. QSs with a QQ phase transition have been studied in recent works~\citep{Zhang:2025sss, Yang:2025hqs}, motivated by several studies of the QCD phase diagram that suggest the existence of multiple distinct phases after deconfinement occurs, such as 2SC, 2SC$+s$, CFL, quark-gluon plasma, LOFF phase~\citep{alford:2001ccc,alford:2001csq,ruester:2005pdo}. 
Even more, within the QQ-HS scenario, we explore the possibility of including (or not) a crust in the outermost layer, motivated by the potential existence of a floating crust above the strange matter core~\citep{Miralda:1990tso, Glendenning:1992nsc}. In the following paragraphs of this section, we describe in detail the models, prescriptions, and parametrizations implemented in each case.

For the hadron sector, we implement a new parametrization of the generalized piecewise polytropic (GPP) EOS originally proposed by~\citet{OBoyle:2020peo}. The GPP EOS is constructed by a sequence of generalized polytropes given by the following: 
\begin{eqnarray}
P_i(\rho) &=& K_i\rho^{\Gamma_i} + \Lambda_i,\\
\varepsilon_i(\rho) &=& \frac{K_i \rho^{\Gamma_i}}{\Gamma_i - 1} + (1 + a_i) \rho - \Lambda_i   ,  \\
c_{s_i}(\rho) &= & \left[\frac{1}{\Gamma_i - 1} + \frac{1 + a_i}{K_i \Gamma_i \rho ^ {\Gamma_i - 1}}\right] ^ {-\frac{1}{2}}   .
\end{eqnarray}
The parameters 
$K_i$, $\Gamma_i$, $a_i$, and $\Lambda_i$ fully characterize the EOS in each density interval $[\rho_{i-1}, \rho_i]$, where the intervals are defined by the parameters $\rho_i$, $i = 1, ..., N$. In our case, we will adopt $N=3$ and the last interval being semi-infinite, $[\rho_2, +\infty)$.  It is important to stress that these parameters are not independent, as the continuity and differentiability of the energy density $\varepsilon(\rho)$ and pressure $p(\rho)$ are enforced at the dividing densities.

\begin{figure}[H]
\includegraphics[width=0.8\linewidth]{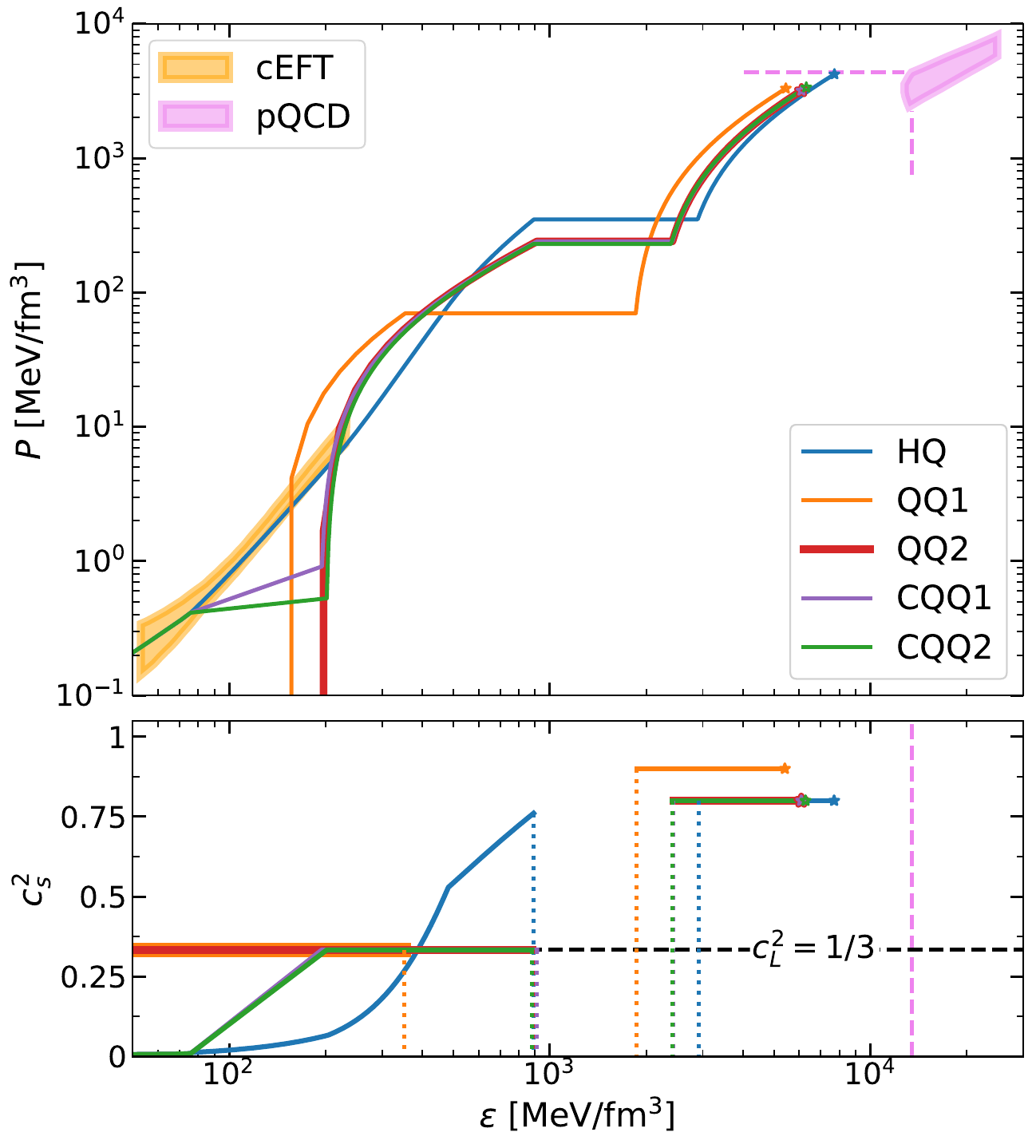}
\caption{$P$-$\varepsilon$ (\textbf{top} panel) 
 and $c_s^2$-$\varepsilon$ (\textbf{bottom} panel) relationships of the hybrid EOSs selected for the CSS ($\beta=0$) analysis. \textls[-12]{In both panels, the star marks represent the maximum energy density value reached}}
\label{fig:peps_beta0}
\end{figure}

This hadronic sector, when applied, is used to model the outer core region with a number density above $0.5~n_0$, where $n_0 = 0.16$~fm$^{-3}$ is the nuclear saturation density; for densities below $0.5~n_0$, we adopt the BPS-BBP crust~\citep{Baym:1971tgs, Baym:1971nsm}. We glue the crust and core EOSs simply concatenating them at $0.5~n_0$, ensuring that the thermodynamic quantities increase monotonically. Within the GPP formalism, the junction density $0.5~n_0$ corresponds to the value $\log_{10}(\rho_0) = 14.127$, $\rho_0$ being only the particular parameter of the GPP method that determines the starting point of the first piece of the GPP EOS. The other details of the particular selected parametrization are presented in Table~\ref{tab:had} and are discussed in the following section along with the results. In addition to the astrophysical motivations discussed later, the BPS-BBP crust and the adopted GPP parametrization have been calibrated to agree with the boundaries of the region permitted by the {$\chi$EFT} constraint reported in~\citet{Drischler:2021lma}, as depicted in the low-density region of Figures~\ref{fig:peps_beta0} and {4}
.

\begin{table}[H] 
\small 
\caption{Parameters 
 of the selected hadronic EOS constructed using the GPP prescription of Ref.~\cite{OBoyle:2020peo}. The adopted value for $\log_{10}(\rho_0)= 14.127$ ensures continuity with the BPS-BBP crust sector at $0.5~n_0$.}
\label{tab:had}
\begin{tabularx}{\textwidth}{c CCCCC CCC}
\toprule
  & \boldmath{$\log_{10}\rho_0$} & \boldmath{$\log_{10}\rho_1$} & \boldmath{$\log_{10}\rho_2$} & \boldmath{$\log_{10}(K_1)$} & \boldmath{$\Gamma_1$} & \boldmath{$\Gamma_2$} & \boldmath{$\Gamma_3$} \\ 
\midrule
Hadronic EOS &  14.127 & 14.55 & 14.90 & $-$27.22 & 2.761 & 3.80 & 2.40 \\
\bottomrule
\end{tabularx}
\end{table}
\vspace{-6pt}
\vspace{-12pt}
{\captionof*{figure}{at the center of the terminal SSHS configuration. Top panel: colored region constraints are provided by $\chi$EFT and pQCD as presented in Ref.~\cite{Drischler:2021lma} and Ref.~\cite{Annala:2020efq}, respectively; the $\chi$EFT constraint is only applicable to hadron and crust matter. The dashed segments near the pQCD constraint indicate the quadrant where dense matter can inhabit before reaching the pQCD regime; as the highest central energy density and pressure for all sets occur within this region, all the selected EOSs are not incompatible with this condition, despite not satisfying $c_s^2 = 1/3$. Bottom panel: each EOS indicates the abrupt phase transition with the empty gap between vertical dashed segments in each curve. The horizontal dashed black line indicates the conformal limit value, $c_L^2 = 1/3$, and the vertical pink dashed line indicates the beginning of pQCD region. After the phase transition all EOSs present a constant speed of sound depending on the respected selected value; although being significantly far from the conformal limit value, all the EOSs reach the terminal energy density before pQCD region.}}
\vspace{+18pt}

For the quark sector, depending on the type of compact object, we use three different models: the MIT Bag, the CSS, and the novel non-CSS one. To model the  quark phase for the outer core in the QQ-HS cases, we use the MIT Bag model, which has only a single free parameter, the bag $B$ that models confinement (see, for example, the review presented in Ref.~\citep{johnson:1975tmb} and references therein). {The EOS of the original version of the MIT bag model is given by
\begin{equation}
    P(\varepsilon)=\frac{1}{3}(\varepsilon - 4B). 
\end{equation}
}

Related to the selection of the traditional MIT model to treat matter in the outer core of QQ-HSs, a comment is in place. Despite the modern, more sophisticated and microscopically involved models of quark matter available in the literature---like the \mbox{(Polyakov-)Nambu--Jona-Lasinio} model (see, for example, Refs.~\cite{buballa:2005nma,orsaria:2014qdi,malfatti:2020dba} and references therein), the Field Correlator Method (see Refs.~\cite{plumari:2013qmi,ranea:2016css,mariani:2017ceh} and references therein), or attempts to present unified EOSs that treat hadrons and quarks within the same Lagrangian (as the Chiral Mean-Field~\cite{Dexheimer:2010ana,Kumar:2024mna} or EVA-01~\cite{celi:2025etr,celi:2025tau-arxiv} models)---they present almost constant speeds of sound similar to those obtained with the simple MIT bag EOS used in this work (see, for example, Refs.~\cite{wang:2019nqs,li:2022son} and references therein). Moreover, the original Bag model has been revisited and modified, including vector interactions and the density dependence for the bag constant (see for example, Refs.~\cite{Lopes:2021mmb,ju:2025hji} and references therein). Additionally, in Ref.~\citep{alford:2005hst} a modification to the MIT EOS that includes phenomenological manner QCD corrections to the pressure of the free-quark Fermi sea and the possibility that quark matter is at a color-flavor-locked color-superconducting phase is presented. We must stress that without the inclusion of vector interactions~\cite{Lopes:2021mmb, malfatti:2020dba}, excluded vacuum effects~\cite{lugones:2024eve} or higher-order quark interactions in the Dirac scalar and vector coupling channels~\cite{benic:2014hhs}, all these models for quark matter tend to have similar low values for the speed of sound, fully compatible with the conformal limit value, $c_L^2=1/3$, but in tension with astronomical observations~\cite{Bedaque:2015svb}. Therefore, macroscopical quantities such as the $M$-$R$ relationships of {QSs} constructed using these different models tend to be morphologically equivalent~\cite{otto:2020haq,Lopes:2021mmb,ju:2025hji}. For these reasons, the original version of the MIT Bag EOS is still used to study effects of quark matter in compact stars (see, for example, Refs.~\cite{rather:2020cbc,mitra:2022hsi,karimi:2023hsw,song:2025csa, Zhang:2025sss}). To sum up, as we aim to explore different compact object proposals in a qualitative and macroscopic manner, for the sake of simplicity, we select the MIT Bag model to model the outer core of the QQ-HSs.

When a crust is also added, as we did for the hadronic sector of the EOS, we use the BPS-BBP model up to $0.5~n_0$ {(and glue it as mentioned above)}; we call these objects Crust QQ (CQQ) stars. For the inner cores of both the HQ and QQ-HS cases, we implement the {CSS} parametrization~\citep{alford:2013gcf}; finally, we also present a novel non-CSS parametric model to describe matter in the inner core of HSs. 

Regarding the CSS model, it has been used to capture the generic behavior of HSs with a sharp first-order {HQ} phase transition through the following three parameters: the phase transition pressure, $P_t$, the energy density gap between phases, $\Delta \varepsilon$, and the squared speed of sound of the quark phase, $c_s^2$, which is assumed to be constant throughout the entire phase. In this model, the EOS is constructed by the low-density model, $\varepsilon_< (P)$, up to $P_t$, and, for larger pressures, the expression for the energy density reads as follows: 
\begin{equation}
   \varepsilon_> (P) = \varepsilon _< (P_t) + \Delta \varepsilon + c_s^{-2} (P-P_t).
   \label{eq:css}
\end{equation}

Regarding the assumption of constant $c_s$, it is important to recall that when SSHSs are taken into account, central energy densities might be as high as those typical of {pQCD} calculations, $n_\textrm{pQCD} \gtrsim 40\, n_0$~\citep{Annala:2020efq}. In particular, it is expected that the squared speed of sound for the quark sector $c_s^2$ approaches the conformal limit, $c_L^2 = 1/3$, for pQCD densities. Consequently, within the SSHS scenario, the CSS parameters should be carefully selected. For this reason, within the CSS model, there exists a limitation in exploring the parameter space when studying SSHSs, so as not to violate the pQCD constraint (the pQCD constraint in the $P$-$\varepsilon$ constraint can be seen in the high density region of Figures~\ref{fig:peps_beta0} and 4). 

With this motivation, the novel non-CSS model proposal is to keep the $P_t$ and $\Delta \varepsilon$ parameters, only modifying the \mbox{$c_s^2 \equiv constant$} condition. We propose a pressure dependent functional form for $c_s^2$ which, for high energy densities, behaves as expected from the pQCD~calculations as follows:
\vspace{-6pt}
\begin{equation} 
    c_s^2(P)=c_L^2+\left( c_t^2 - c_L^2 \right) \exp \left(-\beta  \frac{P-P_t}{P_t}\right) \, .
\end{equation}
The modified parameters are $c_t^2$, the quark speed of sound at the phase transition (which is analogous to $c_s^2$ in the original CSS model), and $\beta$, a dimensionless parameter that regulates the rapidity of the exponential approach to the conformal limit value. For the limiting cases, it values $c_s^2(P_t) = c_t^2$ at the phase transition, and it satisfies the conformal limit for high pressures, $\lim_{P\to\infty} c_s^2(P)=c_L^2$. It is important to notice that when selecting $\beta = 0$, this functional form reduces to the traditional CSS case.

Given that $c_s^2 = \partial P/ \partial \varepsilon$, this proposal is integrable, having an analytical solution for the relationship $P(\varepsilon)$, another key feature of the presented model. Inverting this relationship, and using the same notation as Equation~\eqref{eq:css}, the resulting $\varepsilon_>(P)$ relationship is given by the following:
\begin{equation}
    \varepsilon_> (P) = 
    \varepsilon _< (P_t) + \Delta \varepsilon +
    \frac{P_t}{\beta c_L^{2}} \ln \left\{ 1 + \left(\frac{c_L}{c_t}\right)^2 \left[ \exp\left(\beta \frac{P-P_t}{P_t}\right)-1 \right]   \right\} \, ,
\end{equation}
which, as expected, reduces to the CSS expression given in Equation~\eqref{eq:css} in the $\beta \to 0$ limit. In addition, the proposed parametric functional form for the speed of sound is in agreement with other results presented in works treating quark matter phenomenologically. In Ref.~\cite{ivanytskyi:2022rtc} the conformal limit is reached in a completely similar manner than in this model using the confining density functional approach to treat quark matter. Moreover, in Ref.~\cite{lugones:2024sqs} similar results are obtained within an enhanced version of the quark matter density-dependent model including excluded volume effects. In Ref.~\cite{Traversi:2022sos}, the authors also obtained a $c_s^2$ approaching the conformal limit value from greater values applying the quark EOS model by Ref.~\cite{alford:2005hst} that modifies the MIT Bag model to include CFL effects. Finally, pQCD calculations presented in Refs.~\cite{gorda:2023buq,komoltsev:2024eos} also predict $c_s^2$ approaches the conformal limit from above (see, for example,  Figure~2 from Ref.~\cite{hippert:2025ubo} and the recent review from~\mbox{\citet{kojo:2021qeo}}).

Given all these specific models, we aim to obtain different kinds of HSs. In summary, we plan to construct the following:
\begin{itemize}
    \item HQ-HSs, composed of a BPS-BBP crust, a hadronic GPP outer core detailed in Table~\ref{tab:had}, and an inner core made of quark matter, modeled through both the CCS and non-CSS parametrizations.
    \item QQ-HSs without a crust composed of an outer quark core modeled with the MIT bag model and an inner core of quark matter described using the CSS parametrization.
    \item CQQ-HSs, where the BPS-BBP crust outermost layer is added at $0.5~n_0$ to the previous QQ-HSs configuration.
\end{itemize}

In all cases, as already mentioned, since we aim to explore the implications of the SSHS scenario, we assume an abrupt first-order phase transition in the inner--outer core interface and a slow conversion hypothesis between phases, which gives rise to the SSHS family. {It is important to recall that within the sharp HQ (or QQ) phase transition and the slow conversion regime between phases, stability against radial perturbations does not (necessarily) coincide with the $\partial M/\partial \varepsilon_c > 0$ condition (see, for example, Refs.~\cite{Pereira:2017pte,Lugones:2021pci} and references therein). To address the study of stability against radial perturbations, for each of the presented families of compact objects, we have solved the linearized equations that govern radial perturbations. We have imposed the boundary conditions at the transition radius that allows us to treat consistently the slow conversion hypothesis (details in Ref.~\cite{Pereira:2017pte}). We focus, in particular, on the radial fundamental mode that determines stability. Configurations for which the squared frequency, $\omega_0^2$, is positive (negative) form the (un)stable branches of compact objects.} Using these different proposed configurations {within the slow conversion hypothesis}, we explore the possibility of satisfying the current astrophysical and microphysical constraints.

\section{Results and Discussion} \label{sec:results}

In this section, we present and analyze our results in comparison with other theoretical frameworks. As previously noted, constraints on NSs from $2~M_\odot$ pulsars, gravitational-wave detections, and X-ray observations delineate a scenario that poses significant challenges to contemporary NS models.

Given the different EOSs considered and the stellar configurations proposed (HQ, QQ, and CQQ), we explored the corresponding parameter spaces to satisfy the existing observational constraints. In this initial analysis, the non-CSS model was not considered, and only an inner quark core modeled through the CSS case ($\beta=0$) was included.

Regarding the selected hadronic set, unlike our previous works~\cite{Lugones:2023ama, Mariani:2024cas, Ranea:2023auq, ranea:2022bou, Mariani:2024csi}, where various hadronic parametrizations were examined as qualitative representatives of different families or limiting cases, in this study, we adopt a single GPP parametrization. Although this is a specific choice, it was selected after a thorough exploration of the GPP parameter space considering both microphysical and astrophysical constraints. In this sense, the most recent observational constraints (and their location in the $M$–$R$ plane) are the main reason why we present a single new GPP parametrization rather than applying the multiple enveloping hadronic EOS strategy used in the aforementioned references. This strategy consists of constructing parametric EOSs that act as an envelope for numerous microscopic hadronic models in the literature. In light of the recent observations considered here, however, employing that approach would produce two envelope curves lying very close to each other. Hence, while it would slightly broaden the quantitative analysis of the hadronic EOS, it would not contribute meaningfully to the intended analysis. Moreover, as already noted, we aim to focus on the SSHS hypothesis, and other hadronic EOSs will be introduced when addressing alternative proposals from the literature. In conclusion, for these reasons and for the sake of simplicity, since no relevant feature of the proposal is lost by that, we choose to work with a single hadronic parametrization. The details of the selected hadronic GPP parametrization to model the HQ-HSs are listed in Table~\ref{tab:had}.

On the other hand, regarding quark EOS models, previous works have already implemented them and analyzed in detail the impact of their parameters (the Bag constant and the three CSS ones) on stellar structure and stability. For example, Refs.~\cite{Lopes:2021mmb, Menezes:2005sse} show that a smaller Bag constant \textit{stiffens} the quark EOS, allowing larger radii and masses for QSs. In Ref.~\cite{rather:2020cbc}, the MIT Bag model is applied to construct HQ-HSs, showing that, besides controlling the stiffness of the quark EOS, the Bag constant also modifies the onset of the phase transition as follows: as $B$ increases, the transition occurs at larger pressures and, consequently, the HS branch begins at higher stellar masses. Regarding the CSS model, Refs.~\cite{Christian:2018cot} present the impact of $P_t$ and $\Delta\varepsilon$ on HQ-HSs as follows: $P_t$ sets the transition onset, which shifts to larger stellar masses as $P_t$ increases; $\Delta\varepsilon$ affects the EOS stiffness and thus the slope of the HS branch after the transition, being stiffer for smaller $\Delta\varepsilon$. Ref.~\cite{Ranea:2023auq} reports the same behavior for $\Delta\varepsilon$, and it additionally explores the role of $c_s^2$, finding stiffer EOSs and larger maximum masses for larger $c_s^2$. Moreover, studies on SSHSs showed that, for a fixed hadronic EOS, larger SSHS branches follow from stiffer quark EOSs and, in particular, from larger $\Delta\varepsilon$ values~\cite{Lugones:2023ama, Mariani:2024cas}. In this context, our work does not aim to re-examine the quantitative influence of each parameter but to explore, using these established results, the possibility of satisfying current constraints. For this reason, although we explore the parameter space exhaustively to determine the selected sets, we only present five sets as representative of many qualitatively equivalent cases, focusing instead on the capability of our proposal (relative to other ones in the literature) to account for modern observations. 

In this sense, regarding the MIT Bag model used to construct QQ and CQQ HSs, we select the $B$ values so as to satisfy the maximum mass constrained by \mbox{PSR~J0952-0607}~\cite{Romani:2022pjt} and also to obtain different radii compatible with low-mass constraints (excepting \mbox{XTE~J1814-338} measurement), obtaining values for our sets in the \mbox{$35$~MeV/fm$^3$ $\lesssim B \lesssim$ $50$~MeV/fm$^3$} range. Regarding the CSS model, we made an exhaustive exploration for the three configurations, HQ-, QQ-, and CQQ-HSs, and guided by our previous works~\cite{Lugones:2023ama, Mariani:2024cas}, we explore the ranges,
\begin{eqnarray*}
    10~\mathrm{MeV/fm}^3 \le &P_{t}& \le 400~\mathrm{MeV/fm}^3 \, , \\
    100~\mathrm{MeV/fm}^3 \le &\Delta \epsilon& \le 3000~\mathrm{MeV/fm}^3 \, , \\
    0.33 \le &c_s^2& \le 1.0 \, .
\end{eqnarray*}
We select our sets choosing the $P_t$ value in order to obtain proper maximum masses and the $\Delta \epsilon$ and $c_s^2$ values in order to obtain suitable (long enough) SSHS branches. The obtained values for the selected $P_t$ and $\Delta \epsilon$ are in agreement with previous works that used microphysical models to construct hybrid EOSs as follows: the selected obtained values for $P_t$ are in line with Refs.~\cite{ranea:2016css,ranea:2017csi,tonetto:2020dgm,rather:2024roo} and the selected values for $\Delta\varepsilon$ are compatible with Refs.~\cite{celi:2025etr,celi:2025tau-arxiv,mariani:2019mhs,lenzi:2023hsw}. The values selected for $c_s^2$ are motivated by the related discussion in Section~\ref{sec:intro} and are also discussed later once the results are presented. Based on all these considerations, we have selected five representative cases that serve as suitable scenarios for distinct stellar configurations; the parameters of the these sets are presented in Table~\ref{tab:sets}. In the following, we present and analyze the results for them.

The top panel of Figure~\ref{fig:peps_beta0} shows the $P$-$\varepsilon$ relations for the five selected sets. In each curve, the constant-pressure plateau represents the energy density discontinuity associated with the first-order phase transition, while the small star symbol indicates the maximum energy density and pressure reached in the \textit{terminal mass} configuration, the last stable SSHS. The figure also displays the microphysical constraints derived from $\chi$EFT and pQCD calculations. The pink dashed segments adjacent to the pQCD region roughly delineate the quadrant where NS matter should reside before entering the pQCD regime in order to not be incompatible with it. As mentioned previously, both the crust and the hadronic EOS satisfy the $\chi$EFT constraint, whereas the quark phases are not subject to it. Regarding the pQCD constraint, the SSHSs can reach several tens of the nuclear saturation density in their interiors, with the star symbols lying very close to the pQCD region, although the SSHS matter does not fully enter this regime. Thus, the modeled SSHS matter for these sets---though extreme---remains consistent with pQCD calculations, since the corresponding EOSs terminate within the allowed quadrant. Nevertheless, such extreme behavior highlights the need for a precise parameter adjustment to avoid inconsistency with pQCD, further motivating the consideration of the non-CSS model.

\begin{table}[H] 
\caption{Parametrization 
 details of the two-phase EOSs selected for the CSS ($\beta=0$) analysis; the details regarding the hadron EOS used in the HQ EOS are presented in Table~\ref{tab:had}.} 
\label{tab:sets}
\begin{adjustwidth}{-\extralength}{0cm}
\centering 
\begin{tabularx}{\fulllength}{CCCCCC}
\toprule
\textbf{EOS}	& \textbf{Hadron Sector} & \textbf{Bag [MeV/fm\textsuperscript{3}]} & \boldmath{$P_t$} \textbf{[MeV/fm\textsuperscript{3}]} & \boldmath{$\Delta \varepsilon$} \textbf{[MeV/fm\textsuperscript{3}]} & \boldmath{$c_s^2$} \\
\midrule
HQ		& \cmark & - & 350 & 2000 & 0.8 \\
QQ1		& \xmark & 36 & 70 & 1500 & 0.9 \\
QQ2		& \xmark & 48 & 240 & 1500 & 0.8 \\
CQQ1		& \xmark & 48 & 240 & 1500 & 0.8 \\
CQQ2		& \xmark & 50 & 230 & 1500 & 0.8 \\
\bottomrule
\end{tabularx}
\end{adjustwidth}
\end{table}

The bottom panel of Figure~\ref{fig:peps_beta0} shows the $c_s^2$-$\varepsilon$ relations for the five selected sets. In each curve, the phase transition jump is indicated with an empty gap between vertical dashed segments. There is also displayed the conformal limit value, $c_L^2 = 1/3$, with a horizontal black dashed line, and the vertical pink dashed line marks the onset of the pQCD region. Before the phase transition, each EOS behaves differently as follows: the hadronic EOS presents the usual increase from very low values up to around $c_s^2 \sim 0.7$, and the MIT Bag quark EOS shows the expected $c_s^2 = 1/3$ constant value (except for the low density crust region in the CQQ cases). After the phase transition, by construction, all the EOSs present constant $c_s^2$ behavior; in particular, for the selected EOSs, the $c_s^2$ values are notably outside the conformal limit. However, as already stated in the previous $P$-$\varepsilon$ analysis, since the terminal energy densities (denoted with a star symbol) occur before the pQCD region, these EOSs are not opposed to pQCD predictions.

In Figure~\ref{fig:mraio_beta0}, we present the $M$-$R$ relationships for the five selected sets. Each curve is presented up to the terminal configuration, which is marked with a small star; in this sense, for all sets, the phase transition occurs at the maximum mass configuration, and the long branches after the maximum mass configurations are stable SSHS branches up to the terminal configuration, where stability against radial perturbations is lost. With different color regions, we also present the astrophysical constraints; of particular relevance for this work are the four \textit{extreme} constraints from HESS~J1731-347, PSR~J0952-0607, PSR~J1231-1411, and XTE~J1814-338. As previously discussed, although these measurements are complex and, in some cases, subject to debate, they remain viable and represent a significant challenge for theoretical NS models. In line with the objectives of this work, we selected five representative sets to explore different approaches satisfying the aforementioned observational constraints, or at least a significant subset of them. The HQ curve, in particular, satisfies all constraints, reaching the XTE~J1814-338 region only marginally. In this case, the hadronic GPP parameters were tuned to produce a purely hadronic curve that meets all constraints except for XTE~J1814-338, while the SSHS configuration accounts for this final measurement. {As already discussed, the simultaneous consideration of current constraints (with the exception of XTE~J1814-338) does not leave much room for a qualitatively broad envelope analysis, neither in radius nor in mass, within the HQ-HS proposal, and a single hadronic EOS can thus serve as a suitable representative for this proposal and to address the aimed discussion. The QQ1 case has a low-valued $B$ parameter in the outer core and a very high speed of sound, $c_s^2 = 0.9$ in the inner core. This very stiff EOS set yields large radii and masses for the stellar configurations along the traditional branch, as well as an extended SSHS branch that satisfies the J0614-3329 and XTE~J1814-338 measurements. In contrast, the QQ2 set represents a similar case to QQ1 but with a softer EOS, achieved through a larger $B$ value and a lower speed of sound. This less extreme QQ configuration would be more appropriate if the J1231-1411 measurement is revised in the future. The inclusion of a crust in the CQQ1 and CQQ2 sets allows these models to satisfy the latter constraint without requiring very small $B$ values or excessively stiff EOSs.

\begin{figure}[H]
\includegraphics[width=0.98\linewidth]{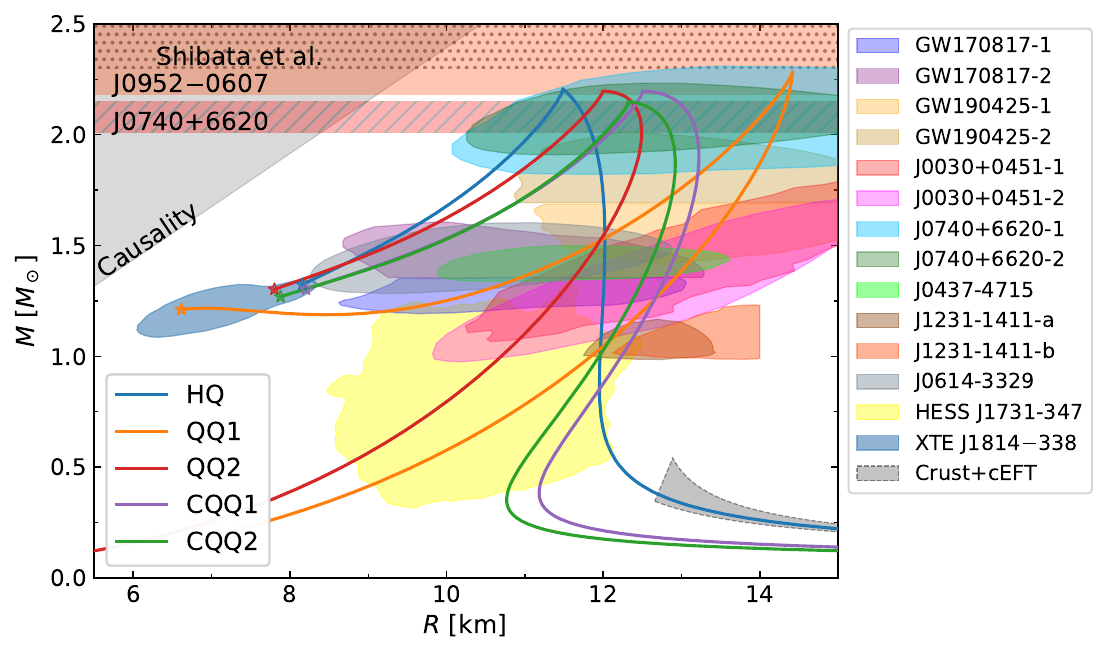}
\caption{$M$-$R$ relationship of the of the hybrid EOSs selected for the CSS ($\beta=0$) analysis. For all sets, we present only stable configurations, considering the slow conversion scenario{; configurations beyond the maximum mass, up to the terminal one (star marks), belong to the SSHS branch.} 
We also show astrophysical constraints from the \mbox{$\sim$2~$M_\odot$} pulsars~\cite{Demorest:2010sdm, Antoniadis:2013amp, Arzoumanian:2018tny, Cromartie:2020rsd, Fonseca:2021rfa}, NICER pulsars~\cite{Miller:2019pjm,Riley2019anv,Miller:2021tro,Riley:2021anv, Salmi:2024anv, Choudhury2024anv,Mauviard:2025anv}, GW170817~\cite{Abbott:2017oog, Abbott:2018gmo} and GW190425~\cite{Abbott:2020goo} events, the \textit{black widow} PSR~J0952-0607~\citep{Romani:2022pjt}, HESS~J1731-347~\cite{Doroshenko:2022asl}, and XTE~J1814-338~\citep{Kini:2024ctp}. The upper horizontal dotted area is the region excluded by~\citet{Shibata:2019ctb}, $M_\textrm{max} \leq 2.3~M_\odot$. The gray area in the right bottom corner indicates the TOV integration of the $\chi$EFT EOS constraint (along with a BPS-BBP crust for lower densities); as already mentioned, this constraint does not apply to compact objects containing quark matter at such masses. The shaded region in the upper left corner indicates the causality forbidden zone.}
\label{fig:mraio_beta0}
\end{figure}

In Figure~\ref{fig:tidal_beta0} we show the dimensionless tidal deformability, $\Lambda$, as a function of the gravitational mass, $M$, for the five selected models constructed within the CSS ($\beta = 0$) case. It is important to remark that CQQ-HSs (that are in agreement with constraints in the $M$-$R$ plane) do not fulfill the restriction coming from the GW170817 event. Moreover, the terminal mass objects of the QQ-HS families have extremely low values of $\Lambda$, completely consistent with the results for black holes that have a null value. All the other selected EOSs satisfy this constraint trough the traditional totally stable branch.

To complement the shown results and provide quantitative details, in Table~\ref{tab:mr14}, we present values for the radius, dimensionless tidal deformability and central energy density of compact objects with $M = 1.4\,M_\odot$, both corresponding to the totally stable branch and the SSHS. In addition, Table~\ref{tab:mrmaxterm} contains the mass, radius, dimensionless tidal deformability, and central energy density of the maximum mass and terminal mass~configurations.

\begin{figure}[H]
\includegraphics[width=0.9\linewidth]{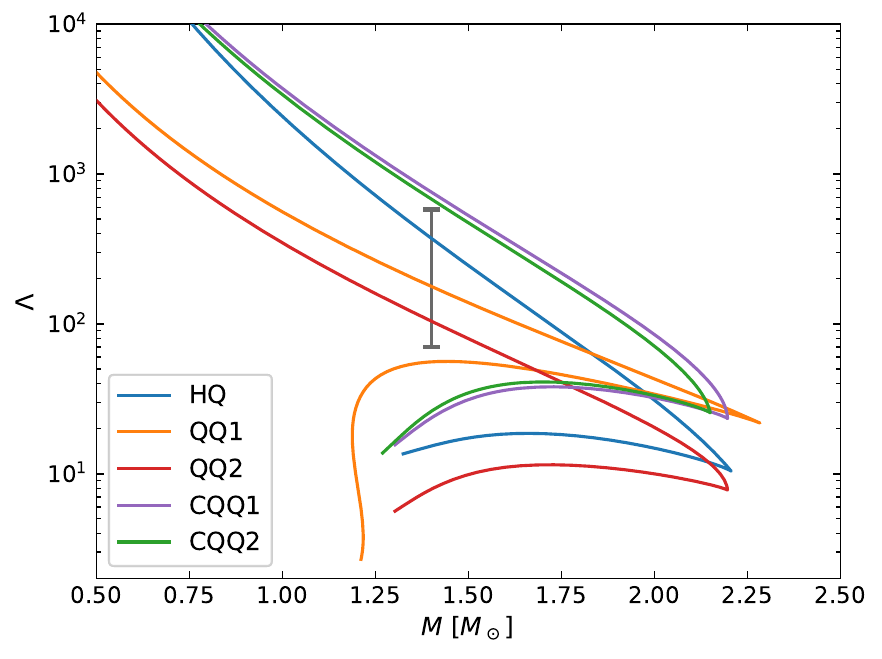}
\caption{{$\Lambda$-$M$ relationship 
 of the of the hybrid EOSs selected for the CSS ($\beta=0$) analysis. For all sets we present only stable configurations, considering the slow conversion scenario{; configurations beyond the maximum mass, up to the terminal one at the end of the curve, belong to the SSHS branch.} 
We also show the constraint obtained from the analysis of the multimessenger event with gravitational waves GW170817~\cite{Abbott:2018gmo}. The CQQ cases, although satisfying the $M$-$R$ constraint for GW170817 (see Figure~\ref{fig:mraio_beta0}), does not satisfy the $\Lambda$ one. All the other selected EOSs satisfy it trough the totally stable~branch.}}
\label{fig:tidal_beta0}
\end{figure}

Considering all the $P$-$\varepsilon$, $c_s^2$-$\varepsilon$, $M$-$R$, and $\Lambda$-$M$ analyses, and despite the limitations in evaluating the CSS parameters within the SSHS and pQCD frameworks, all the selected cases indicate that the SSHS scenario is a promising candidate for simultaneously satisfying the current astrophysical and microphysical constraints. Although a precise tuning of the model parameters is required, our results show that more than one family of HSs can yield consistent outcomes. Within the slow-conversion regime, we obtain models of HQ-, QQ-, and CQQ-HSs capable of addressing the present observational tensions {coming from microphysics and in the $M$-$R$ plane. Besides these constraints, the HQ and QQ cases also satisfy the $\Lambda$ constraint, while the CQQ cases do not. In these particular CQQ cases, the technique used to glue the crust--core interface should be revised in order to also satisfy the existing $\Lambda$ constraint (see Refs.~\cite{Fortin:2016nsr, Canullan:2025ccc} to see details regarding different crust--core interpolation methods and their impact on the stellar structure). On the other hand, a} common feature among all studied cases is the necessity of adopting a very high speed of sound, $c_s^2 \sim$ 0.8--0.9. As already comprehensively discussed in Section~\ref{sec:eos}, although such values are difficult to achieve in many effective microphysical quark EOS models, repulsive interactions, as vector interactions, excluded vacuum effects or higher-order quark interactions in the Dirac scalar and vector coupling channels can substantially increase the stiffness. Nevertheless, as previously discussed, the conformal limit predicted by pQCD at extremely high densities imposes a stringent upper bound on this quantity.

In addition to the CSS cases analyzed, we also investigate the novel non-CSS scenario. For this purpose, as we aimed to study the impact of the non-CSS parametrization, we consider a specific HQ-HS configuration that, within the CSS framework ($\beta = 0$), satisfies the constraints in the $M$-$R$ plane but fails to meet the pQCD requirement. For this scenario, we only present HQ-HS configurations, and not the QQ and CQQ alternatives, in order to avoid redundancy, since the three alternatives share the same qualitative results regarding the non-CSS effects. In this case, we take, by construction, a particularly extreme set to effectively test the non-CSS proposal as follows:
\begin{align*}
    P_t &= 400 \ \textrm{MeV/fm}^3 \, , \\
    \Delta \varepsilon &= 2500 \ \textrm{MeV/fm}^3 \, , \\
    c_t^2 &= 1.0 \, .
\end{align*}
Given this set, we explore different EOSs by varying the $\beta$ parameter. For more details on the numerical values of relevant magnitudes, see Table~\ref{tab:mrbetavar} of Appendix~\ref{app:A}. We present the corresponding results for the $P$-$\varepsilon$ (Figure~\ref{fig:peps_betavar}, top panel), $c_s^2$-$\varepsilon$ (Figure~\ref{fig:peps_betavar}, bottom panel), and $M$-$R$ (Figure~\ref{fig:mraio_betavar}) planes. In the \mbox{$P$-$\varepsilon$} diagram, one can observe that increasing the $\beta$ value gradually softens the EOS, and it eventually brings it into agreement with the pQCD constraints. In the particular case presented, with $\beta=0.1$, the pQCD constraint is already (marginally) satisfied, while the $\beta=0.4, 0.7$ satisfies it much more naturally. The $c_s^2$-$\varepsilon$ diagram also evidences this feature as follows: after the phase transition, it can be seen how the speed of sound decreases exponentially from the initial selected value, $c_t^2 = 1$ in the current case, towards the conformal limit value. In this panel, the $\beta=0.4, 0.7$ cases are also distinguished for rapidly approaching this limit value. Beyond the specific details, it can be observed that, in general, the implementation of the non-CSS approach helps to alleviate the tension with the pQCD regime in the SSHS scenario. However, the $M$-$R$ plane evidences the shortcomings of this novel method as follows: the results in that plane show that increasing $\beta$ strongly shortens the SSHS branch, making it difficult or even impossible to reach the most extreme XTE~J1814-338 measurement. Therefore, although this novel model provides a promising framework for satisfying the conformal limit of the speed of sound while maintaining stiffness in the lower-density regions of the quark EOS, its inherent limitations appear unavoidable as long as the current extreme observational constraints hold. Table~\ref{tab:mrbetavar} presents values for the mass, radius, dimensionless tidal deformability, and central energy density of compact objects with $M = 1.4\,M_\odot$, corresponding to the totally stable branch and the SSHS branch, and of the maximum mass and terminal mass configurations obtained for the non-CCS~EOSs.

In what follows, we would like to discuss and compare the proposals and results of this work (along with the previous SSHS results presented in the preceding works by~\citet{Lugones:2023ama} and~\citet{Mariani:2024cas}) in the context of other recent theoretical proposals existing in the literature (already mentioned in Section~\ref{sec:intro}). The proposals discussed in this section are not meant to form an exhaustive survey, but rather to provide a qualitative representation of the existing theoretical models. Firstly, there exists the so-called \textit{two-family} scenario, proposed originally by~\citet{Drago:2014cvc} and also developed in more recent works~\citep{Drago:2016tso1, Drago:2016tso2, DiClemente:2024itc}. In this proposal, a hadron NS family and a self-bound QS family would coexist, where the NSs would explain the low-mass--low-radius objects and the QSs the high-radius--high-mass objects. The recent work by~\citet{Shirke:2025pan} also points in the direction of a strange QS family favored by the recent PSR~J0614-3329 observation. A comprehensive list of alternatives is presented in the work of~\citet{Sagun:2023wit}, which explores the potential explanations for the measurement of HESS~J1731-347. In this work, the authors mentioned the possibility of including different exotic degrees of freedom inside NSs, such as an early HQ phase transition or a DM contribution. On the other side, the recent work by~\citet{Pal:2025cah} studied totally stable HQ-HSs, including both early and late HQ phase transitions, considering only NICER observations. Finally, along with the most recent NS detections from XTE~J1814-338 and HESS~J1731-347, there appear a couple of works that also aim to explain current observations in different SSHS scenarios as follows:~\citet{Laskos:2025xja} implemented HQ-HSs and~\citet{Zhang:2025sss} proposed the existence of self-bound QQ-HSs.

\begin{figure}[H]
\includegraphics[width=0.7\linewidth]{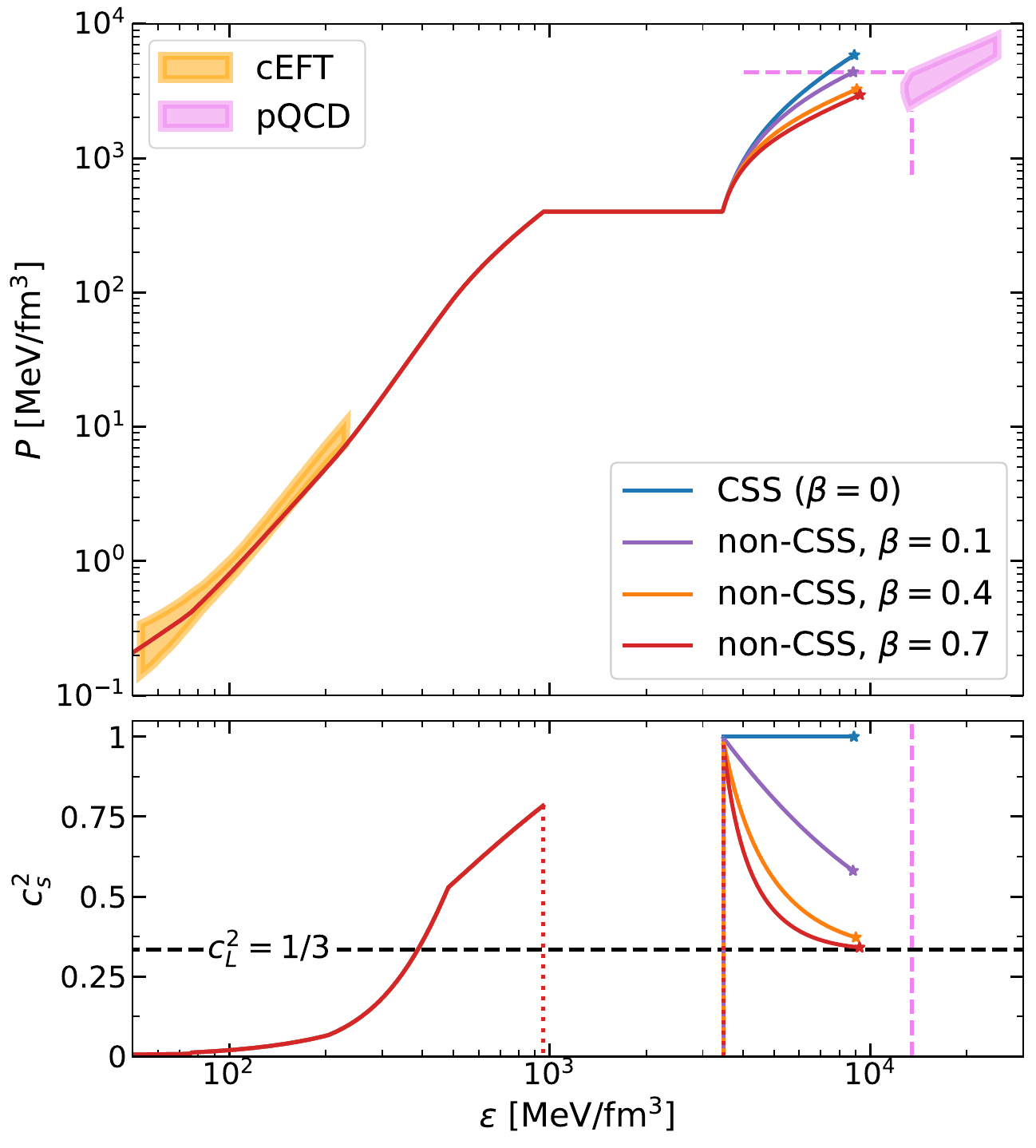}
\caption{$P$-$\varepsilon$ (\textbf{top} panel) 
 and $c_s^2$-$\varepsilon$ (\textbf{bottom} panel) relationships of the hybrid EOSs selected for the non-CSS ($\beta \neq 0$) analysis. In both panels, the star marks represent the maximum energy density value reached at the center of the terminal SSHS configuration. Top panel: colored regions and dashed segment details are in caption of Figure~\ref{fig:peps_beta0}. Except for the $\beta$ parameter (detailed in the legend for each curve), the shared parameters for all curves are $P_t = 400$~MeV/fm$^3$, $\Delta \varepsilon = 2500$~MeV/fm$^3$, $c_t^2 = 1.0$. The increasing $\beta$ value helps to satisfy the pQCD constraint. {Bottom panel: each EOS indicates the abrupt phase transition with the empty gap between vertical dashed segments in each curve. The horizontal dashed black line indicates the conformal limit value, $c_L^2 = 1/3$, and the vertical pink dashed line indicates the beginning of the pQCD region. After the phase transition all EOSs begins with same value $c_s^2 = 1.0$ and, depending on the $\beta$ value, decrease exponentially with different rapidity approaching the conformal limit. The $\beta=0.4$ and $\beta=0.7$ cases decrease enough to almost reach this limit well before the beginning of the pQCD region.}}
\label{fig:peps_betavar}
\end{figure}

In Figure~\ref{fig:mraio_literature}, we present the $M$-$R$ relationships for some of the existing proposals available in the mentioned literature. All the curves presented in the figure are obtained from the articles mentioned in the last paragraph. In this sense, it is important to note that the specific curves chosen to illustrate each proposal are not intended to be fully representative of the complete set of results reported in the corresponding references. Rather, they are selected to provide a graphical illustration of each proposal and to facilitate a qualitative assessment of its overall shape, as well as its potential strengths and limitations. In this framework, we seek to compare and discuss these various proposals in light of our results and the current astronomical constraints.

\begin{figure}[H]
\includegraphics[width=0.98\linewidth]{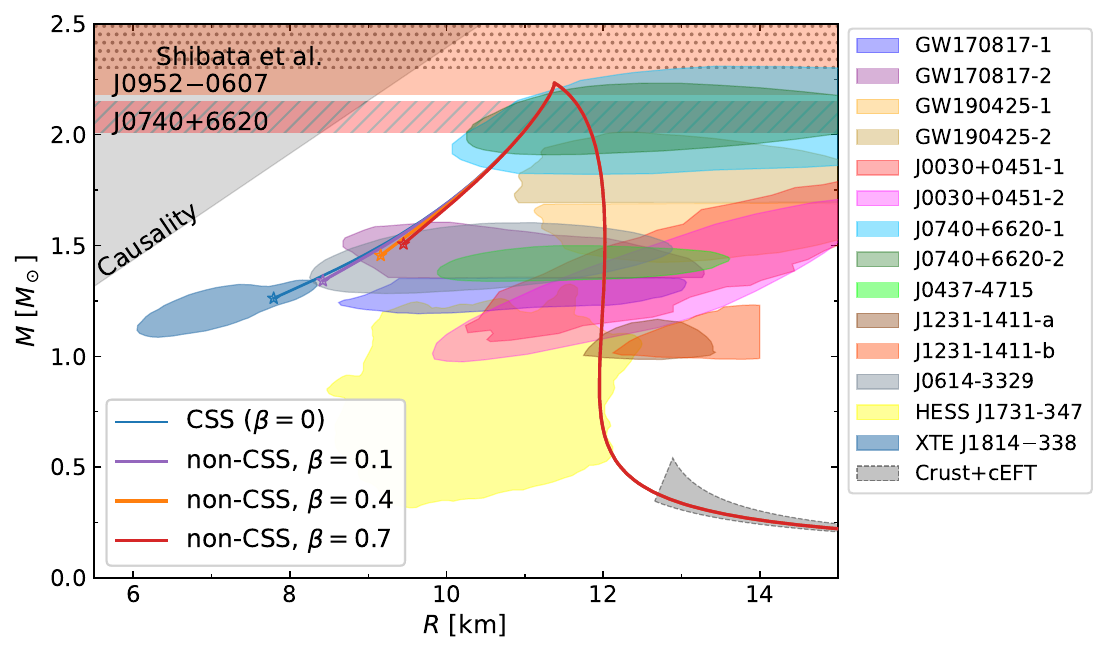}
\caption{$M$-$R$ relationship of the of the hybrid EOSs selected for the non-CSS ($\beta \neq 0$) analysis. For all sets we present only stable configurations, considering the slow conversion scenario, these being the configurations after the maximum mass one up to the terminal one (star marks) of the SSHS branch. Colored regions constraints are detailed in Figure~\ref{fig:mraio_beta0}.  Except for the $\beta$ parameter (detailed in the legend for each curve), the shared parameters for all curves are $P_t = 400$~MeV/fm$^3$, $\Delta \varepsilon = 2500$~MeV/fm$^3$, and $c_t^2 = 1.0$. The increasing $\beta$ value, while helping satisfy the pQCD constraint, reduces the possibility of meeting the XTE~J1814-338 constraint.}
\label{fig:mraio_betavar}
\end{figure}

Firstly, the two-family scenario, presented with the blue and cyan pair of curves, appears to be promising since the hadron NS family (blue curve) could be adjusted to also meet the XTE~J1814-338 constraint, and the QS family (cyan curve) is able to reach most of the other measurements; however, this proposal does not seem to be fully capable of satisfying the J1231-1411 NICER observation without missing other constraints. The~\mbox{\citet{Sagun:2023wit}} proposals---comprising an HQ-HS configuration with an early phase transition (orange curve) and an admixed DM-NS model (red curve)---can, through suitable parameter adjustments, reproduce HESS~J1731-347 and J1231-1411, but they fail to simultaneously explain XTE~J1814-338. The~\citet{Pal:2025cah} model of fully stable HQ-HSs with both early (olive curve) and late (dark blue curve) phase transition displays similar strengths and limitations to those of the~\citet{Sagun:2023wit} framework. The~\mbox{\citet{Laskos:2025xja}} (green curve when considering the APR hadronic EOS and purple curve for DD2 one) and~\citet{Zhang:2025sss} models (brown curve) serve as precursors to the results presented here; however, they are not tuned to reproduce the extreme measurements of J1231-1411 and J0952-0607. Furthermore, both of these earlier works exhibit the need for extreme $c_s^2$ values; this previously discussed potential difficulty of the SSHS models should be further verified and analyzed in these works to assess the viability of their proposals. In all the cases presented, although we do not show the corresponding $P$-$\varepsilon$ relations for these models, the fact that none of the hadronic proposals we considered satisfy the {$\chi$EFT} $M$-$R$ region suggests that their respective EOSs likely fail to meet this constraint. While different {$\chi$EFT} calculations exist---and thus the particular {$\chi$EFT} constraint shown here is not the only one available---we would like to emphasize that a suitable compact star proposal should, at the very least, take one of them into account. Finally, the constraint by~\citet{Shibata:2019ctb}, $M_\textrm{max} \leq 2.3~M_\odot$, becomes relevant in this context; there exists a narrow viable mass range between the J0952-0607 measurement and this maximum mass constraint that may warrant some consideration when proposing models. On the other hand, any of the mentioned works present results for the speed of sound of their EOSs and only the works by~\citet{Pal:2025cah} and~\citet{Zhang:2025sss} present results for the dimensionless tidal deformability. Therefore, only a brief analysis can be made with respect to these quantities.~\citet{Pal:2025cah} show that the HQ-HSs with a very early phase transition satisfy this constraint trough the hybrid branch, and the late phase transition HQ-HSs satisfy that trough hadronic branches; \mbox{\citet{Zhang:2025sss}} show that, within their QQ-HS model, the constraint can be meet only trough the SSHS branch. Although this particular analysis regarding these compact objects families is model-dependent, it is important to remark that in any theoretical proposal the speed of sound and the dimensionless tidal deformability are relevant quantities that should be~verified.

\begin{figure}[H]
\includegraphics[width=0.97\linewidth]{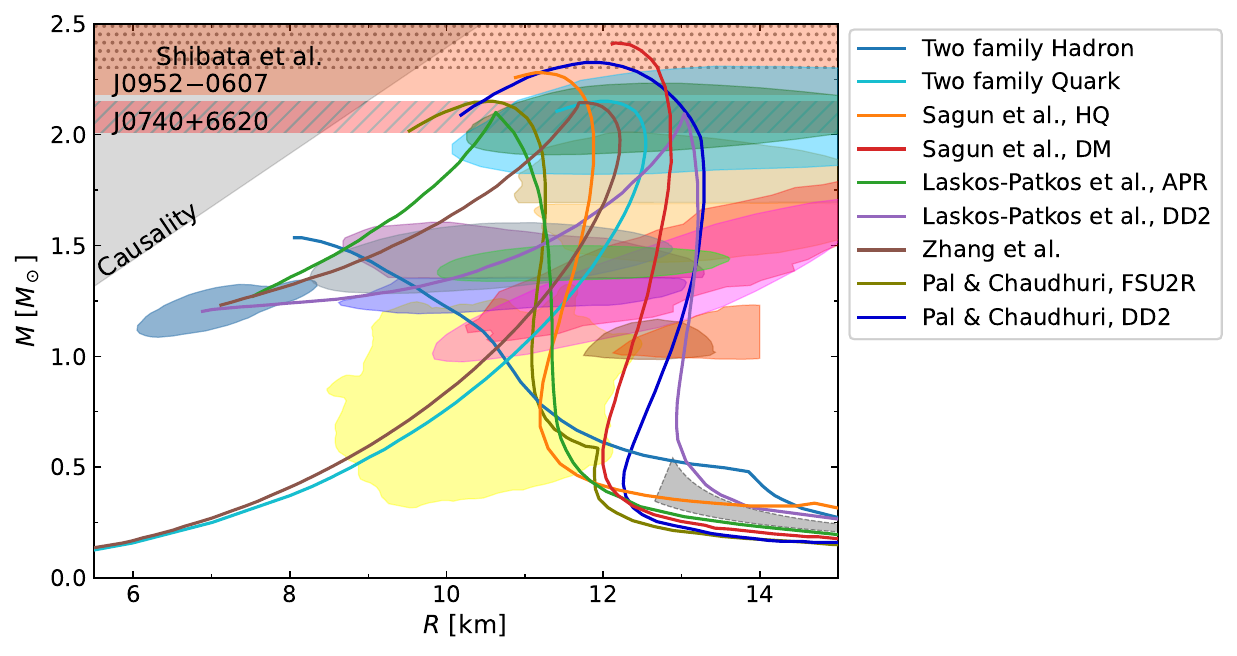}
\caption{$M$-$R$ relationship of different existing proposals of the literature. Colored regions constraints are detailed in Figure~\ref{fig:mraio_beta0}. Details regarding the selected models and the corresponding references are presented in the main text.}
\label{fig:mraio_literature}
\end{figure}

\section{Conclusions and Future Perspectives} \label{sec:conclusion}

In this work we aim to explain the current astrophysical and microphysical constraints on NSs invoking the SSHS scenario in a model-independent manner. We construct, within this hypothesis, different kinds of compact objects, as the HQ-, QQ-, and CQQ-HSs, and, exploring the parameter space, we select some representative sets capable of satisfying the mentioned constraints coming from microphysics and in the $M$-$R$ plane. Besides these constraints, we also analyzed the dimensionless tidal deformability constraint; only the CQQ cases do not satisfy it due to the crust--core interpolation method used, which should be revised in future studies. We compare our results with other existing NS theoretical proposals of the literature, aiming to review the current state-of-the-art in this area.

Despite obtaining satisfactory results within our models, they show the tension and zero-sum game that the modern astrophysical and microphysical constraints configure currently. In this context, the SSHS proposal, even considering its shortcomings, remains as one of the promising strategies to explain and describe the measured compact objects. However, it should also be said that some of the \textit{extreme} detections, \mbox{HESS~J1731-347}, \mbox{PSR~J0952-0607}, PSR~J1231-1411, and XTE~J1814-338, generate some skepticism due to the potential impact of the hypothesis used in each estimation process, and any rectification over its measured values should imply a revision of these conclusions.

To summarize our findings, we present a takeaway of the major conclusions obtained in this work as follows:
\begin{itemize}
    \item The modern picture of astronomical constraints related to compact objects produces strong tensions, and models with some type of exotic matter seem to be favored. In particular, if the current estimations of either XTE~J1814-338 or PSR~J1231-1411 are confirmed (or not strongly rectified) by future analysis, the need to include some type of exotic matter in the inner core of compact stars might be strongly favored.
    
    \item  \textls[-25]{In accordance to previous proposals presented by~\citet{Lugones:2023ama} and~\mbox{\citet{Mariani:2024cas}}} (and also in line with alternative scenarios recently presented in Refs.~\citep{Laskos:2025xja, Zhang:2025sss}), we show that SSHSs could lead to an appropriate description of modern astronomical observations of compact objects (even considering the extreme ones). Despite this being true, large values of the speed of sound are needed, and potential issues with the conformal limit of pQCD might arise for long SSHS branches.
    
    \item \textls[-15]{Contrary to the traditional CSS model, the novel non-CSS parametrization proposed in this work is useful for avoiding potential problems with pQCD calculations, but it introduces issues, particularly when explaining the challenging XTE~J1814-338~observation.}
    
    \item The analysis of other recent proposals from the literature---including regular hadronic NSs, the two-family scenario, admixed DM HSs and NSs, and QQ-HSs---shows that, while all leave some room for further refinement or updating, none of them is entirely suitable. Whether considered jointly or separately, XTE~J1814-338 and PSR~J1231-1411 place stringent constraints on these models. 
    
    \item Despite these limitations, the other proposed hadronic and hybrid models are in tension with $\chi$EFT calculations. If these ideas are used in the future, the low-pressure 
    region needs to be adjusted.
\end{itemize}

To close this work, we would like to emphasize some potentially relevant observational aspects where attention must be directed in order to shed some light into the behavior of matter under extreme conditions of pressure.

Future observational signaling of a $g$-mode associated with a sharp first-order phase transition might provide strong evidence to favor the existence of SSHSs as those non-radial oscillation modes are only excited in the slow conversion scenario~\citep{tonetto:2020dgm}. For this reason, this might be the most promising observational aspect to discern between purely hadronic objects and hybrid ones; moreover, it could shed some light into the nature of the HQ phase transition~\citep{sotani:2001ddo,miniutti:2003nro,ranea:2018omo,rodriguez:2021hsw}. Another quantity to which attention must be paid is the dimensionless tidal deformability, as it has been proven to show extremely different behavior for NSs and SSHSs of a given gravitational mass. This is particularly evident for values $M \lesssim 1.6\,M_\odot$ (see, for example, Refs.~\citep{Lugones:2023ama,ranea:2022bou,Ranea:2023auq,ranea:2023cmr,Pal:2025cah}). Upcoming detections of GWs emitted by isolated (proto-)NSs are expected to provide an even more direct probe of the internal composition of these compact objects. Gravitational-wave asteroseismology can reveal characteristic oscillation modes whose frequencies and damping times depend very sensitively on the internal structure and composition of NSs. Observing these signals could therefore yield decisive evidence regarding the occurrence, nature, and properties of a HQ phase transition in the depths of these compact objects.

The next generation of observational facilities promises to revolutionize our understanding of compact stars and the behavior of ultra-dense matter. The continuing operation and planned upgrades of gravitational-wave detectors, along with the forthcoming third-generation observatories, will enable the detection of a larger population of binary NS mergers and the follow up of post-merger remnants with unprecedented sensitivity (see, for example, Ref.~\cite{Luck:2020tgg} and references therein). Simultaneously, high-precision X-ray timing observations will refine measurements of NSs masses and radii (and potentially also restrict their moments of inertia), providing tighter constraints on the dense matter EOS (see, for example, Refs.~\cite{li:2025dmi,Majczyna:2020pom} and references therein).  Together, these multi-messenger observations are expected to deliver decisive empirical evidence regarding the highest densities known in the universe and the existence and nature of HQ phase transitions.

\vspace{6pt}

\authorcontributions{M.M. and I.F.R.-S. 
 contributed equally to the conceptualization, formal analysis, investigation, visualization, writing, and every other aspect of work presented in this paper. All authors have read and agreed to the published version of the manuscript.}

\funding{This research was funded by CONICET and UNLP (Argentina) under grants PIP 0169, G187, and G009.
}

\dataavailability{The original contributions presented in this study are included in the article. Further inquiries can be directed to the corresponding author.
}

\acknowledgments{The authors would like to thank the anonymous referees for the constructive criticisms that help improve the quality of the work.
}

\conflictsofinterest{The authors declare no conflicts of interest.
} 





\appendixtitles{no} 
\appendixstart
\appendix 
\section[\appendixname~\thesection]{} \label{app:A}

In this appendix, we present quantitative data for magnitudes of relevant stellar configurations we calculate, present, and analyze in the main body of the article. Although the current work has a qualitative approach, we present here the following data in order to provide more details of our model and to contribute to the reproducibility and clarity of our study.

In Table~\ref{tab:mr14} we present the radius, dimensionless tidal deformability, and central energy density for the $M=1.4 M_\odot$ stellar configurations for the five selected hybrid EOSs within the CSS ($\beta=0$) analysis; in Table~\ref{tab:mrmaxterm} we present the mass, radius, dimensionless tidal deformability, and central energy density of the maximum mass and the SSHS terminal configuration for the same five EOSs. In Table~\ref{tab:mrbetavar} we present the same data for the relevant stellar configurations obtained in the non-CSS analysis.

\begin{table}[H]
\small
\caption{Properties of $M=1.4 M_\odot$ stellar configurations for each hybrid EOSs for the CSS ($\beta=0$) analysis. Within the slow conversion scenario, we obtain two \textit{twin} $1.4 M_\odot$ configurations for each EOS, the traditional totally stable one (before maximum mass) and the SSHS configuration one (after maximum mass). For each one, we present the radius $R_{1.4}$, dimensionless tidal deformability $\Lambda_{1.4}$, and central energy density $\varepsilon_{c,1.4}$.}
\label{tab:mr14}

\begin{adjustwidth}{-\extralength}{0cm}
\centering 
\begin{tabularx}{\fulllength}{CCCCCCC}
\toprule
\multirow{3.5}{*}{\textbf{EOS}} & \multicolumn{3}{c}{\textbf{Totally Stable} \boldmath{$M=1.4 M_\odot$} \textbf{Configuration}} & \multicolumn{3}{c}{\textbf{SSHS} \boldmath{$M=1.4 M_\odot$} \textbf{Configuration}} \\
\cmidrule{2-7} 
& \boldmath{$R_{1.4}$} & \boldmath{$\Lambda_{1.4}$} & \boldmath{$\varepsilon_{c,1.4}$} & \boldmath{$R_{1.4}$} & \boldmath{$\Lambda_{1.4}$} & \boldmath{$\varepsilon_{c,1.4}$} \\
& \textbf{{[}km{]}} &  & \textbf{{[}MeV/fm\textsuperscript{3}{]}} & \textbf{{[}km{]}} & & \textbf{{[}MeV/fm\textsuperscript{3}{]}} \\
\midrule
HQ   & 12.02 & 373.0 & 462.0 & 8.60 & 15.3 & 6338.0 \\
QQ1  & 13.04 & 177.7 & 214.0 & 11.26 & 55.7 & 2198.5 \\
QQ2  & 11.72 & 104.4 & 309.7 & 8.58 & 7.9 & 4620.3 \\
CQQ1 & 12.85 & 758.3 & 309.5 & 9.03 & 23.6 & 4622.7 \\
CQQ2 & 12.57 & 680.7 & 327.0 & 9.06 & 26.1 & 4269.8 \\
\bottomrule
\end{tabularx}
\end{adjustwidth}
\end{table}
\vspace{-9pt}

\begin{table}[H]
\caption{Properties of maximum mass and terminal (last stable SSHS) stellar configurations for each hybrid EOSs for the CSS ($\beta=0$) analysis. For each one, we present mass $M$, radius $R$, dimensionless tidal deformability $\Lambda$, and central energy density $\varepsilon_{c}$.}
\label{tab:mrmaxterm}

\begin{adjustwidth}{-\extralength}{0cm}
\centering 
\begin{tabularx}{\fulllength}{CCCCCCCCC}
\toprule
\multirow{3.5}{*}{\textbf{EOS}} & \multicolumn{4}{c}{\textbf{Maximum Mass Configuration}} & \multicolumn{4}{c}{\textbf{SSHS Terminal Configuration}} \\
\cmidrule{2-9} 
& \boldmath{$M_\textrm{\textbf{max}}$} & \boldmath{$R_{\textrm{\textbf{max}}}$} & \boldmath{$\Lambda_{\textrm{\textbf{max}}}$} & \boldmath{$\varepsilon_{c,\textrm{\textbf{max}}}$} & \boldmath{$M_\textrm{\textbf{term}}$} & \boldmath{$R_{\textrm{\textbf{term}}}$} & \boldmath{$\Lambda_{\textrm{\textbf{term}}}$} & \boldmath{$\varepsilon_{c,\textrm{\textbf{term}}}$} \\
& \boldmath{{[}$M_\odot${]}} & \textbf{{[}km{]}} &  & \textbf{{[}MeV/fm\textsuperscript{3}{]}} & \boldmath{{[}$M_\odot${]}} & \textbf{{[}km{]}} & & \textbf{{[}MeV/fm\textsuperscript{3}{]}} \\
\midrule
HQ   & 2.21 & 11.48 & 10.4 & 2892.2 & 1.32 & 8.16 & 13.6 & 7722.5 \\
QQ1  & 2.28 & 14.42 & 21.8 & 1854.0 & 1.21 & 6.61 & 2.7 & 5447.4 \\
\bottomrule
\end{tabularx}
\end{adjustwidth}
\end{table}
\begin{table}[H]\ContinuedFloat
\caption{\textit{Cont.}}
\begin{adjustwidth}{-\extralength}{0cm}
\centering 
\begin{tabularx}{\fulllength}{CCCCCCCCC}
\toprule
\multirow{3.5}{*}{\textbf{EOS}} & \multicolumn{4}{c}{\textbf{Maximum Mass Configuration}} & \multicolumn{4}{c}{\textbf{SSHS Terminal Configuration}} \\
\cmidrule{2-9} 
& \boldmath{$M_\textrm{\textbf{max}}$} & \boldmath{$R_{\textrm{\textbf{max}}}$} & \boldmath{$\Lambda_{\textrm{\textbf{max}}}$} & \boldmath{$\varepsilon_{c,\textrm{\textbf{max}}}$} & \boldmath{$M_\textrm{\textbf{term}}$} & \boldmath{$R_{\textrm{\textbf{term}}}$} & \boldmath{$\Lambda_{\textrm{\textbf{term}}}$} & \boldmath{$\varepsilon_{c,\textrm{\textbf{term}}}$} \\
& \boldmath{{[}$M_\odot${]}} & \textbf{{[}km{]}} &  & \textbf{{[}MeV/fm\textsuperscript{3}{]}} & \boldmath{{[}$M_\odot${]}} & \textbf{{[}km{]}} & & \textbf{{[}MeV/fm\textsuperscript{3}{]}} \\
\midrule
QQ2  & 2.20 & 12.00 & 7.8 & 2412.0 & 1.30 & 7.80 & 5.6 & 6094.5 \\
CQQ1 & 2.20 & 12.50 & 23.3 & 2412.0 & 1.30 & 8.20 & 15.6 & 6094.5 \\
CQQ2 & 2.15 & 12.32 & 25.5 & 2390.0 & 1.27 & 7.88 & 13.8 & 6310.2 \\
\bottomrule
\end{tabularx}
\end{adjustwidth}
\end{table}
\vspace{-8pt}
\begin{table}[H]
\caption{Properties of SSHS $M=1.4 M_\odot$ and the terminal stellar configurations for each hybrid EOS for the \mbox{non-CSS ($\beta \ne 0$)} analysis; the $\beta=0.4$ and $\beta=0.7$ cases do not present SSHS \mbox{$M=1.4 M_\odot$} configurations. For each one, we present the mass $M$ (for terminal configuration only), radius $R$, dimensionless tidal deformability $\Lambda$, and central energy density $\varepsilon_c$. All selected non-CSS EOSs share the same properties for the totally stable $M=1.4 M_\odot$ ($R_{1.4} = 12.02$~km, \mbox{$\Lambda_{1.4} = 373.0$,} \mbox{$\varepsilon_{c,1.4} = 462.0$~MeV/fm$^3$),} and maximum mass ($M_\textrm{max} = 2.23 M_\odot$, $R_{\textrm{max}} = 11.37$~km, $\Lambda_{\textrm{max}}=8.5$, $\varepsilon_{c,\textrm{max}}= 3456.8$~MeV/fm$^3$) configurations.}
\label{tab:mrbetavar}

\begin{adjustwidth}{-\extralength}{0cm}
\centering 
\begin{tabularx}{\fulllength}{CCCCCCCC}
\toprule
\multirow{3.5}{*}{\textbf{EOS}} & \multicolumn{3}{c}{\textbf{SSHS} \boldmath{$M=1.4 M_\odot$} \textbf{Configuration}} & \multicolumn{4}{c}{\textbf{SSHS Terminal Configuration}} \\
\cmidrule{2-8} 
& \boldmath{$R_{1.4}$} & \boldmath{$\Lambda_{1.4}$} & \boldmath{$\varepsilon_{c,1.4}$} & \boldmath{$M_\textrm{\textbf{term}}$} & \boldmath{$R_{\textrm{\textbf{term}}}$} & \boldmath{$\Lambda_{\textrm{\textbf{term}}}$} &$\boldmath\varepsilon_{c,\textrm{\textbf{term}}}$ \\
& \textbf{{[}km{]}} &  & \textbf{{[}MeV/fm\textsuperscript{3}{]}} & \boldmath{{[}$M_\odot${]}} & \textbf{{[}km{]}} & & \textbf{{[}MeV/fm\textsuperscript{3}{]}} \\
\midrule
\mbox{$\beta=0$ (CSS)}& 8.63 & 15.8 & 6444.3 & 1.26 & 7.79 & 12.3 & 8911.1 \\
$\beta =0.1$ & 8.72 & 17.3 & 7412.2 & 1.34 & 8.42 & 16.8 & 8836.5 \\
$\beta =0.4$ & - & - & - & 1.45 & 9.16 & 21.2 & 9065.9 \\
$\beta =0.7$ & - & - & - & 1.51 & 9.45 & 22.6 & 9282.5 \\
\bottomrule
\end{tabularx}
\end{adjustwidth}
\end{table}


\begin{adjustwidth}{-\extralength}{0cm}

\reftitle{References}



\PublishersNote{}
\end{adjustwidth}

\begin{thebibliography}{999}
\bibitem[{Hewish} et~al.(1968){Hewish}, {Bell}, {Pilkington}, {Scott}, and
{Collins}]{Hewish:1968ofa}
{{Hewish},} 
A.; {Bell}, S.J.; {Pilkington}, J.D.H.; {Scott}, P.F.; {Collins},
R.A.
\newblock {Observation of a Rapidly Pulsating Radio Source}.
\newblock {\em Nature} {\bf 1968}, {\em 217},~709--713. [\href{http://dx.doi.org/10.1038/217709a0}{CrossRef}]
\bibitem[{\"O}zel and Freire(2016)]{ozel:2016mra}
{\"O}zel, F.; Freire, P.
\newblock Masses, radii, and the equation of state of neutron stars.
\newblock {\em Annu. Rev. Astron. Astrophys.} {\bf 2016}, {\em 54},~401--440. [\href{http://dx.doi.org/10.1146/annurev-astro-081915-023322}{CrossRef}]
\bibitem[Vidana(2018)]{vidana:2018asw}
Vidana, I.
\newblock A short walk through the physics of neutron stars.
\newblock {\em Eur. Phys. J. Plus} {\bf 2018}, {\em 133},~445. [\href{http://dx.doi.org/10.1140/epjp/i2018-12329-x}{CrossRef}]
\bibitem[Kondratyev(2019)]{kondratyev:2019mom}
Kondratyev, V.
\newblock Magnetoemission of Magnetars.
\newblock {\em Phys. Part. Nucl.} {\bf 2019}, {\em 50},~613--615. [\href{http://dx.doi.org/10.1134/S1063779619050150}{CrossRef}]
\bibitem[Esposito et~al.(2020)Esposito, Rea, and Israel]{esposito:2020mas}
Esposito, P.; Rea, N.; Israel, G.L.
\newblock Magnetars: A short review and some sparse considerations.
\newblock In {\em Timing Neutron Stars: Pulsations, Oscillations and Explosions}; {Springer:  Berlin/Heidelberg, Germany}, 2020; pp. 97--142.
\bibitem[Demorest et~al.(2010)Demorest, Pennucci, Ransom, Roberts, and
Hessels]{Demorest:2010sdm}
Demorest, P.; Pennucci, T.; Ransom, S.; Roberts, M.; Hessels, J.
\newblock {Shapiro Delay Measurement of A Two Solar Mass Neutron Star}.
\newblock {\em Nature} {\bf 2010}, \emph{{467}}, 1081--{1083}. [\href{http://dx.doi.org/10.1038/nature09466}{CrossRef}]
\bibitem[Antoniadis et~al.(2013)]{Antoniadis:2013amp}
Antoniadis, J.; Freire, P.C.C.; Wex, N.; Tauris, T.M.; Lynch, R.S.; van Kerkwijk, M.H.; Kramer, M.; Bassa, C.; Dhillon, V.S.; Driebe, T.; et al.
\newblock {A Massive Pulsar in a Compact Relativistic Binary}.
\newblock {\em Science} {\bf 2013}, \emph{{340}}, 6131. [\href{http://dx.doi.org/10.1126/science.1233232}{CrossRef}]
\bibitem[Arzoumanian et~al.(2018)]{Arzoumanian:2018tny}
Arzoumanian, Z.; Brazier, A.; Burke-Spolaor, S.; Chamberlin, S.; Chatterjee, S.; Christy, B.; Cordes, J.M.; Cornish, N.J.; Crawford, F.; Cromartie, H.T.; et al.
\newblock The {NANOGrav} 11-year Data Set: High-precision Timing of 45
Millisecond Pulsars.
\newblock {\em Astrophys. J. Suppl. Ser.} {\bf 2018}, {\em 235},~37. [\href{http://dx.doi.org/10.3847/1538-4365/aab5b0}{CrossRef}]
\bibitem[{Cromartie} et~al.(2020)]{Cromartie:2020rsd}
{Cromartie}, H.T.; Fonseca, E.; Ransom, S.M.; Demorest, P.B.; Arzoumanian, Z.; Blumer, H.; Brook, P.R.; DeCesar, M.E.; Dolch, T.; Ellis, J.A.; et al.
\newblock {Relativistic Shapiro delay measurements of an extremely massive
millisecond pulsar}.
\newblock {\em Nat. Astron.} {\bf 2020}, {\em 4},~72--76. [\href{http://dx.doi.org/10.1038/s41550-019-0880-2}{CrossRef}]
\bibitem[{Fonseca} et~al.(2021)]{Fonseca:2021rfa}
{Fonseca}, E.; Cromartie, H.T.; Pennucci, T.T.; Ray, P.S.; Kirichenko, A.Y.; Ransom, S.M.; Demorest, P.B.; Stairs, I.H.; Arzoumanian, Z.; Guillemot, L.; et al.
\newblock {Refined Mass and Geometric Measurements of the High-mass PSR
J0740+6620}.
\newblock {\em    Astrophys. J. Lett.} {\bf 2021}, {\em 915},~L12. [\href{http://dx.doi.org/10.3847/2041-8213/ac03b8}{CrossRef}]
\bibitem[Miller and et~al.(2019)]{Miller:2019pjm}
Miller, M.C.; Lamb, F.K.; Dittmann, A.J.; Bogdanov, S.; Arzoumanian, Z.; Gendreau, K.C.; Guillot, S.; Harding, A.K.; Ho, W.C.G.; Lattimer, J.M.; et al.
\newblock {PSR} J0030+0451 Mass and Radius from {NICER} Data and Implications
for the Properties of Neutron Star Matter.
\newblock {\em Astrophys. J. Lett.} {\bf 2019}, {\em 887},~L24. [\href{http://dx.doi.org/10.3847/2041-8213/ab50c5}{CrossRef}]
\bibitem[Riley and et~al.(2019)]{Riley2019anv}
Riley, T.E.; Watts, A.L.; Bogdanov, S.; Ray, P.S.; Ludlam, R.M.; Guillot, S.; Arzoumanian, Z.; Baker, C.L.; Bilous, A.V.; Chakrabarty, D.; et al.
\newblock A {NICER} View of {PSR} J0030+0451: Millisecond Pulsar Parameter
Estimation.
\newblock {\em Astrophys. J. Lett.} {\bf 2019}, {\em 887},~L21. [\href{http://dx.doi.org/10.3847/2041-8213/ab481c}{CrossRef}]
\bibitem[{Miller} et~al.(2021)]{Miller:2021tro}
{Miller}, M.C.; Lamb, F.K.; Dittmann, A.J.; Bogdanov, S.; Arzoumanian, Z.; Gendreau, K.C.; Guillot, S.; Ho, W.C.G.; Lattimer, J.M.; Loewenstein, M.; et al.
\newblock {The Radius of PSR J0740+6620 from NICER and XMM-Newton Data}.
\newblock {\em Astrophys. J. Lett.} {\bf 2021}, {\em 918},~L28. [\href{http://dx.doi.org/10.3847/2041-8213/ac089b}{CrossRef}]
\bibitem[{Riley} et~al.(2021)]{Riley:2021anv}
{Riley}, T.E.; Watts, A.L.; Ray, P.S.; Bogdanov, S.; Guillot, S.; Morsink, S.M.; Bilous, A.V.; Arzoumanian, Z.; Choudhury, D.; Deneva, J.S.; et al.
\newblock {A NICER View of the Massive Pulsar PSR J0740+6620 Informed by Radio
Timing and XMM-Newton Spectroscopy}.
\newblock {\em Astrophys. J. Lett.} {\bf 2021}, {\em 918},~L27. [\href{http://dx.doi.org/10.3847/2041-8213/ac0a81}{CrossRef}]
\bibitem[{Salmi} et~al.(2024){Salmi}, {Deneva}, {Ray}, {Watts}, {Choudhury},
{Kini}, {Vinciguerra}, {Cromartie}, {Wolff}, {Arzoumanian}, {Bogdanov},
{Gendreau}, {Guillot}, {Ho}, {Morsink}, {Cognard}, {Guillemot}, {Theureau},
and {Kerr}]{Salmi:2024anv}
{Salmi}, T.; {Deneva}, J.S.; {Ray}, P.S.; {Watts}, A.L.; {Choudhury}, D.;
{Kini}, Y.; {Vinciguerra}, S.; {Cromartie}, H.T.; {Wolff}, M.T.;
{Arzoumanian}, Z.;  et~al.
\newblock {A NICER View of PSR J1231‑1411: A Complex Case}.
\newblock {\em Astrophys. J. Lett.} {\bf 2024}, {\em 976},~58. [\href{http://dx.doi.org/10.3847/1538-4357/ad81d2}{CrossRef}]
\bibitem[{Choudhury} et~al.(2024){Choudhury}, {Salmi}, {Vinciguerra}, {Riley},
{Kini}, {Watts}, {Dorsman}, {Bogdanov}, {Guillot}, {Ray}, {Reardon},
{Remillard}, {Bilous}, {Huppenkothen}, {Lattimer}, {Rutherford},
{Arzoumanian}, {Gendreau}, {Morsink}, and {Ho}]{Choudhury2024anv}
{Choudhury}, D.; {Salmi}, T.; {Vinciguerra}, S.; {Riley}, T.E.; {Kini}, Y.;
{Watts}, A.L.; {Dorsman}, B.; {Bogdanov}, S.; {Guillot}, S.; {Ray}, P.S.;
et~al.
\newblock {A NICER View of the Nearest and Brightest Millisecond Pulsar: PSR
J0437{\textendash}4715}.
\newblock {\em Astrophys. J. Lett.} {\bf 2024}, {\em 971},~L20. [\href{http://dx.doi.org/10.3847/2041-8213/ad5a6f}{CrossRef}]
\bibitem[{Mauviard} et~al.(2025){Mauviard}, {Guillot}, {Salmi}, {Choudhury},
{Dorsman}, {Gonz{\'a}lez-Caniulef}, {Hoogkamer}, {Huppenkothen}, {Kazantsev},
{Kini}, {Olive}, {Stammler}, {Watts}, {Mendes}, {Rutherford}, {Schwenk},
{Svensson}, {Bogdanov}, {Kerr}, {Ray}, {Guillemot}, {Cognard}, and
{Theureau}]{Mauviard:2025anv}
{Mauviard}, L.; {Guillot}, S.; {Salmi}, T.; {Choudhury}, D.; {Dorsman}, B.;
{Gonz{\'a}lez-Caniulef}, D.; {Hoogkamer}, M.; {Huppenkothen}, D.;
{Kazantsev}, C.; {Kini}, Y.;  et~al.
\newblock {A NICER view of the 1.4 solar-mass edge-on pulsar PSR J0614--3329}.
\newblock {\em arXiv} {\bf 2025}, arXiv:2506.14883. [\href{http://dx.doi.org/10.48550/arXiv.2506.14883}{CrossRef}]
\bibitem[Abbott et~al.(2017)]{Abbott:2017oog}
Abbott, B.P.; Abbott, R.; Abbott, T.D.; Acernese, F.; Ackley, K.; Adams, C.; Adams, T.; Addesso, P.; Adhikari, R.X.; Adya, V.B.; et al.
\newblock GW170817: Observation of Gravitational Waves from a Binary Neutron
Star Inspiral.
\newblock {\em Phys. Rev. Lett.} {\bf 2017}, {\em 119},~161101. [\href{http://dx.doi.org/10.1103/PhysRevLett.119.161101}{CrossRef}] [\href{http://www.ncbi.nlm.nih.gov/pubmed/29099225}{PubMed}]
\bibitem[Abbott et~al.(2018)]{Abbott:2018gmo}
{Abbott, B.; Abbott, R.; Abbott, T.D.; Acernese, F.; Ackley, K.; Adams, C.; Adams, T.; Addesso, P.; Adhikari, R.X.; Adya, V.B.; et al.
GW170817: Measurements of neutron star radii and equation of state.} 
\newblock {\em Phys. Rev. Lett.} {\bf 2018}, \emph{{121}}, 161101. [\href{http://dx.doi.org/10.1103/PhysRevLett.121.161101}{CrossRef}] [\href{http://www.ncbi.nlm.nih.gov/pubmed/30387654}{PubMed}]
\bibitem[{Abbott} et~al.(2020)]{Abbott:2020goo}
{Abbott}, B.P.;  Abbott, R.; Abbott, T.D.; Abraham, S.; Acernese, F.; Ackley, K.; Ackley, C.; Adams, R.X.; Adhikari, V.B.; Adya, C.; et~al.
\newblock {GW190425: Observation of a Compact Binary Coalescence with Total
Mass {\ensuremath{\sim}} 3.4 M$_{{\ensuremath{\odot}}}$}.
\newblock {\em Astrophys. J. Lett.} {\bf 2020}, {\em 892},~L3. [\href{http://dx.doi.org/10.3847/2041-8213/ab75f5}{CrossRef}]
\bibitem[{Romani} et~al.(2022){Romani}, {Kandel}, {Filippenko}, {Brink}, and
{Zheng}]{Romani:2022pjt}
{Romani}, R.W.; {Kandel}, D.; {Filippenko}, A.V.; {Brink}, T.G.; {Zheng}, W.
\newblock {PSR J0952-0607: The Fastest and Heaviest Known Galactic Neutron
Star}.
\newblock {\em Astrophys. J. Lett.} {\bf 2022}, {\em 934},~L17. [\href{http://dx.doi.org/10.3847/2041-8213/ac8007}{CrossRef}]
\bibitem[{Doroshenko} et~al.(2022){Doroshenko}, {Suleimanov}, {P{\"u}hlhofer},
and {Santangelo}]{Doroshenko:2022asl}
{Doroshenko}, V.; {Suleimanov}, V.; {P{\"u}hlhofer}, G.; {Santangelo}, A.
\newblock {A strangely light neutron star within a supernova remnant}.
\newblock {\em Nat. Astron.} {\bf 2022}, {\em 6},~1444--1451. [\href{http://dx.doi.org/10.1038/s41550-022-01800-1}{CrossRef}]
\bibitem[{Kini} et~al.(2024){Kini}, {Salmi}, {Vinciguerra}, {Watts}, {Bilous},
{Galloway}, {van der Wateren}, {Khalsa}, {Bogdanov}, {Buchner}, and
et~al.]{Kini:2024ctp}
{Kini}, Y.; {Salmi}, T.; {Vinciguerra}, S.; {Watts}, A.L.; {Bilous}, A.;
{Galloway}, D.K.; {van der Wateren}, E.; {Khalsa}, G.P.; {Bogdanov}, S.;
{Buchner}, J.;  et~al.
\newblock {Constraining the properties of the thermonuclear burst oscillation
source XTE J1814-338 through pulse profile modelling}.
\newblock {\em MNRAS} {\bf 2024}, {\em 535},~1507--1525. [\href{http://dx.doi.org/10.1093/mnras/stae2398}{CrossRef}]
\bibitem[Drischler et~al.(2021)Drischler, Han, Lattimer, Prakash, Reddy, and
Zhao]{Drischler:2021lma}
Drischler, C.; Han, S.; Lattimer, J.M.; Prakash, M.; Reddy, S.; Zhao, T.
\newblock Limiting masses and radii of neutron stars and their implications.
\newblock {\em Phys. Rev. C} {\bf 2021}, {\em 103},~045808. [\href{http://dx.doi.org/10.1103/PhysRevC.103.045808}{CrossRef}]
\bibitem[{Annala} et~al.(2020){Annala}, {Gorda}, {Kurkela}, {N{\"a}ttil{\"a}},
and {Vuorinen}]{Annala:2020efq}
{Annala}, E.; {Gorda}, T.; {Kurkela}, A.; {N{\"a}ttil{\"a}}, J.; {Vuorinen}, A.
\newblock {Evidence for quark-matter cores in massive neutron stars}.
\newblock {\em Nat. Phys.} {\bf 2020}, {\em 16},~907--910. [\href{http://dx.doi.org/10.1038/s41567-020-0914-9}{CrossRef}]
\bibitem[Alford and Halpern(2023)]{alford:2023dcc}
Alford, J.; Halpern, J.
\newblock Do central compact objects have carbon atmospheres?
\newblock {\em Astrophys. J.} {\bf 2023}, {\em 944},~36. [\href{http://dx.doi.org/10.3847/1538-4357/acaf55}{CrossRef}]
\bibitem[Malik et~al.(2025)Malik, Cartaxo, and Provid{\^e}ncia]{Malik:2025oco}
Malik, T.; Cartaxo, J.; Provid{\^e}ncia, C.
\newblock {Observational constraints on neutron star matter equation of state}.
\newblock {\em J. Subat. Part. Cosmol.} {\bf 2025}, \emph{{4}}, 100086. [\href{http://dx.doi.org/10.1016/j.jspc.2025.100086}{CrossRef}]
\bibitem[{Punturo} et~al.(2010){Punturo}, {Abernathy}, {Acernese}, {Allen},
{Andersson}, {Arun}, {Barone}, {Barr}, {Barsuglia}, {Beker}, {Beveridge},
{Birindelli}, {Bose}, {Bosi}, {Braccini}, {Bradaschia}, {Bulik}, {Calloni},
{Cella}, {Chassande Mottin}, {Chelkowski}, {Chincarini}, {Clark}, {Coccia},
{Colacino}, {Colas}, {Cumming}, {Cunningham}, {Cuoco}, {Danilishin},
{Danzmann}, {De Luca}, {De Salvo}, {Dent}, {De Rosa}, {Di Fiore}, {Di
Virgilio}, {Doets}, {Fafone}, {Falferi}, {Flaminio}, {Franc}, {Frasconi},
{Freise}, {Fulda}, {Gair}, {Gemme}, {Gennai}, {Giazotto}, {Glampedakis},
{Granata}, {Grote}, {Guidi}, {Hammond}, {Hannam}, {Harms}, {Heinert},
{Hendry}, {Heng}, {Hennes}, {Hild}, {Hough}, {Husa}, {Huttner}, {Jones},
{Khalili}, {Kokeyama}, {Kokkotas}, {Krishnan}, {Lorenzini}, {L{\"u}ck},
{Majorana}, {Mandel}, {Mandic}, {Martin}, {Michel}, {Minenkov}, {Morgado},
{Mosca}, {Mours}, {M{\"u}ller{\textendash}Ebhardt}, {Murray}, {Nawrodt},
{Nelson}, {Oshaughnessy}, {Ott}, {Palomba}, {Paoli}, {Parguez},
{Pasqualetti}, {Passaquieti}, {Passuello}, {Pinard}, {Poggiani}, {Popolizio},
{Prato}, {Puppo}, {Rabeling}, {Rapagnani}, {Read}, {Regimbau}, {Rehbein},
{Reid}, {Rezzolla}, {Ricci}, {Richard}, {Rocchi}, {Rowan}, {R{\"u}diger},
{Sassolas}, {Sathyaprakash}, {Schnabel}, {Schwarz}, {Seidel}, {Sintes},
{Somiya}, {Speirits}, {Strain}, {Strigin}, {Sutton}, {Tarabrin},
{Th{\"u}ring}, {van den Brand}, {van Leewen}, {van Veggel}, {van den Broeck},
{Vecchio}, {Veitch}, {Vetrano}, {Vicere}, {Vyatchanin}, {Willke}, {Woan},
{Wolfango}, and {Yamamoto}]{punturo:2010tet}
{Punturo}, M.; {Abernathy}, M.; {Acernese}, F.; {Allen}, B.; {Andersson}, N.;
{Arun}, K.; {Barone}, F.; {Barr}, B.; {Barsuglia}, M.; {Beker}, M.;  et~al.
\newblock {The Einstein Telescope: A third-generation gravitational wave
observatory}.
\newblock {\em Class. Quantum Gravity} {\bf 2010}, {\em 27},~194002. [\href{http://dx.doi.org/10.1088/0264-9381/27/19/194002}{CrossRef}]
\bibitem[{Reitze} et~al.(2019){Reitze}, {Adhikari}, {Ballmer}, {Barish},
{Barsotti}, {Billingsley}, {Brown}, {Chen}, {Coyne}, {Eisenstein}, {Evans},
{Fritschel}, {Hall}, {Lazzarini}, {Lovelace}, {Read}, {Sathyaprakash},
{Shoemaker}, {Smith}, {Torrie}, {Vitale}, {Weiss}, {Wipf}, and
{Zucker}]{reitze:2019cet}
{Reitze}, D.; {Adhikari}, R.X.; {Ballmer}, S.; {Barish}, B.; {Barsotti}, L.;
{Billingsley}, G.; {Brown}, D.A.; {Chen}, Y.; {Coyne}, D.; {Eisenstein}, R.;
et~al.
\newblock {Cosmic Explorer: The U.S. Contribution to Gravitational-Wave
Astronomy beyond LIGO}.
\newblock \emph{{arXiv}
}  \textbf{2019}, {arXiv:1907.04833.}
\newblock [\href{http://dx.doi.org/10.48550/arXiv.1907.04833}{CrossRef}]
\bibitem[L{\"u}ck et~al.(2020)L{\"u}ck, Smith, and Punturo]{Luck:2020tgg}
L{\"u}ck, H.; Smith, J.; Punturo, M., Third-Generation Gravitational-Wave
Observatories.
\newblock In {\em Handbook of Gravitational Wave Astronomy}; Bambi, C.,
Katsanevas, S., Kokkotas, K.D., Eds.; Springer: Singapore,  2020;
pp. 1--18. [\href{http://dx.doi.org/10.1007/978-981-15-4702-7_7-1}{CrossRef}]
\bibitem[Bombaci(2017)]{bombaci:2017thp}
Bombaci, I.
\newblock The hyperon puzzle in neutron stars.
\newblock In Proceedings of the 12th International
Conference on Hypernuclear and Strange Particle Physics (HYP2015), {Sendai, Japan, 7--12 September 2015}; p. 101002. 
\bibitem[{Ye} et~al.(2025){Ye}, {Wang}, {Wang}, and {Chen}]{ye:2025hds}
{Ye}, J.T.; {Wang}, R.; {Wang}, S.P.; {Chen}, L.W.
\newblock {High-density Symmetry Energy: A Key to the Solution of the Hyperon
Puzzle}.
\newblock {\em Astrophys. J. Lett.} {\bf 2025}, {\em 985},~238. [\href{http://dx.doi.org/10.3847/1538-4357/add017}{CrossRef}]
\bibitem[{Halasz} et~al.(1998){Halasz}, {Jackson}, {Shrock}, {Stephanov}, and
{Verbaarschot}]{Halasz:1998pfd}
{Halasz}, M.A.; {Jackson}, A.D.; {Shrock}, R.E.; {Stephanov}, M.A.;
{Verbaarschot}, J.J.M.
\newblock {Phase diagram of QCD}.
\newblock {\em Phys. Rev. D} {\bf 1998}, {\em 58},~096007. [\href{http://dx.doi.org/10.1103/PhysRevD.58.096007}{CrossRef}]
\bibitem[{Stephanov}(2006)]{Stephanov:2006qpd}
{Stephanov}, M.
\newblock {QCD phase diagram: An overview}.
\newblock In Proceedings of the XXIVth International Symposium on Lattice Field
Theory,  {Tucson, AZ, USA, 23--28} December 2006; p. 24.1. [\href{http://dx.doi.org/10.22323/1.032.0024}{CrossRef}]
\bibitem[{Guenther}(2020)]{Guenther:2020oot}
{Guenther}, J.N.
\newblock {Overview of the QCD phase diagram---Recent progress from the
lattice}.
\newblock {\em arXiv} {\bf 2020}, arXiv:2010.15503. [\href{http://dx.doi.org/10.48550/arXiv.2010.15503}{CrossRef}]
\bibitem[{Chen} et~al.(2024){Chen}, {Dong}, {He}, {Huang}, {Liu}, {Luo}, {Ma},
{Ruan}, {Shao}, {Shi}, {Sun}, {Tang}, {Tang}, {Wang}, {Wang}, {Wang}, {Xiao},
{Xie}, {Xu}, {Xu}, {Xu}, {Yang}, {Yang}, {Zha}, {Zhang}, {Zhang}, {Zhao}, and
{Zhu}]{Chen:2024pot}
{Chen}, J.H.; {Dong}, X.; {He}, X.H.; {Huang}, H.Z.; {Liu}, F.; {Luo}, X.F.;
{Ma}, Y.G.; {Ruan}, L.J.; {Shao}, M.; {Shi}, S.S.;  et~al.
\newblock {Properties of the QCD matter: Review of selected results from the
relativistic heavy ion collider beam energy scan (RHIC BES) program}.
\newblock {\em Nucl. Sci. Tech.} {\bf 2024}, {\em 35},~214. [\href{http://dx.doi.org/10.1007/s41365-024-01591-2}{CrossRef}]
\bibitem[{Lu} et~al.(2025){Lu}, {Gao}, and {Liu}]{Lu:2025rtf}
{Lu}, Y.; {Gao}, F.; {Liu}, Y.x.
\newblock {Revisiting the first-order QCD phase transition in dense strong
interaction matter}.
\newblock {\em arXiv} {\bf 2025}, arXiv:2509.02974. [\href{http://dx.doi.org/10.48550/arXiv.2509.02974}{CrossRef}]
\bibitem[{Glendenning} and {Weber}(1992)]{Glendenning:1992nsc}
{Glendenning}, N.K.; {Weber}, F.
\newblock {Nuclear Solid Crust on Rotating Strange Quark Stars}.
\newblock {\em Astrophys. J. Lett.} {\bf 1992}, {\em 400},~647. [\href{http://dx.doi.org/10.1086/172026}{CrossRef}]
\bibitem[{Heiselberg} and {Hjorth-Jensen}(1999)]{Heiselberg:1999pti}
{Heiselberg}, H.; {Hjorth-Jensen}, M.
\newblock {Phase Transitions in Neutron Stars and Maximum Masses}.
\newblock {\em Astrophys. J. Lett.} {\bf 1999}, {\em 525},~L45--L48. [\href{http://dx.doi.org/10.1086/312321}{CrossRef}]
\bibitem[{Reddy} et~al.(2000){Reddy}, {Bertsch}, and {Prakash}]{Reddy:2000fop}
{Reddy}, S.; {Bertsch}, G.; {Prakash}, M.
\newblock {First order phase transitions in neutron star matter: Droplets and
coherent neutrino scattering}.
\newblock {\em Phys. Lett. B} {\bf 2000}, {\em 475},~1--8. [\href{http://dx.doi.org/10.1016/S0370-2693(00)00049-6}{CrossRef}]
\bibitem[{Chamel} et~al.(2013){Chamel}, {Fantina}, {Pearson}, and
{Goriely}]{Chamel:2013pti}
{Chamel}, N.; {Fantina}, A.F.; {Pearson}, J.M.; {Goriely}, S.
\newblock {Phase transitions in dense matter and the maximum mass of neutron
stars}.
\newblock {\em Astron. Astrophys.} {\bf 2013}, {\em 553},~A22. [\href{http://dx.doi.org/10.1051/0004-6361/201220986}{CrossRef}]
\bibitem[{Komoltsev}(2024)]{Komoltsev:2024fop}
{Komoltsev}, O.
\newblock {First-order phase transitions in the cores of neutron stars}.
\newblock {\em Phys. Rev. D} {\bf 2024}, {\em 110},~L071502. [\href{http://dx.doi.org/10.1103/PhysRevD.110.L071502}{CrossRef}]
\bibitem[Sch\"afer and Wilczek(1999)]{Schafer:1999coq}
Sch\"afer, T.; Wilczek, F.
\newblock Continuity of Quark and Hadron Matter.
\newblock {\em Phys. Rev. Lett.} {\bf 1999}, {\em 82},~3956--3959. [\href{http://dx.doi.org/10.1103/PhysRevLett.82.3956}{CrossRef}]
\bibitem[Masuda et~al.(2013)Masuda, Hatsuda, and Takatsuka]{Masuda:2013hqc}
Masuda, K.; Hatsuda, T.; Takatsuka, T.
\newblock Hadron–quark crossover and massive hybrid stars.
\newblock {\em Prog. Theor. Exp. Phys.} {\bf 2013}, {\em 2013}, {073D01}. [\href{http://dx.doi.org/10.1093/ptep/ptt045}{CrossRef}]
\bibitem[Hirono and Tanizaki(2019)]{Hirono:2019qhc}
Hirono, Y.; Tanizaki, Y.
\newblock Quark-Hadron Continuity beyond the Ginzburg-Landau Paradigm.
\newblock {\em Phys. Rev. Lett.} {\bf 2019}, {\em 122},~212001. [\href{http://dx.doi.org/10.1103/PhysRevLett.122.212001}{CrossRef}]
\bibitem[{Brandes} et~al.(2023){Brandes}, {Weise}, and
{Kaiser}]{Brandes:2023eaa}
{Brandes}, L.; {Weise}, W.; {Kaiser}, N.
\newblock {Evidence against a strong first-order phase transition in neutron
star cores: Impact of new data}.
\newblock {\em Phys. Rev. D} {\bf 2023}, {\em 108},~094014. [\href{http://dx.doi.org/10.1103/physrevd.108.094014}{CrossRef}]
\bibitem[Fujimoto et~al.(2025)Fujimoto, Fukushima, Hotokezaka, and
Kyutoku]{Fujimoto:2025shq}
Fujimoto, Y.; Fukushima, K.; Hotokezaka, K.; Kyutoku, K.
\newblock Signature of hadron-quark crossover in binary-neutron-star mergers.
\newblock {\em Phys. Rev. D} {\bf 2025}, {\em 111},~063054. [\href{http://dx.doi.org/10.1103/PhysRevD.111.063054}{CrossRef}]
\bibitem[{Fukushima}(2025)]{Fukushima2025qpd}
{Fukushima}, K.
\newblock {QCD Phase Diagram and Astrophysical Implications}.
\newblock {\em arXiv} {\bf 2025}, arXiv:2501.01907. [\href{http://dx.doi.org/10.1016/j.jspc.2025.100066}{CrossRef}]
\bibitem[{Fujimoto} et~al.(2025){Fujimoto}, {Fukushima}, {Hidaka}, and
{McLerran}]{Fujimoto:2025nso}
{Fujimoto}, Y.; {Fukushima}, K.; {Hidaka}, Y.; {McLerran}, L.
\newblock {New state of matter between the hadronic phase and the quark-gluon
plasma?}
\newblock {\em Phys. Rev. D} {\bf 2025}, {\em 112},~074006. [\href{http://dx.doi.org/10.1103/h71y-km92}{CrossRef}]
\bibitem[{Baym, Gordon and Hatsuda, Tetsuo and Kojo, Toru and Powell, Philip D.
and Song, Yifan and Takatsuka, Tatsuyuki}(2018)]{Baym:2018fht}
{Baym, G.; Hatsuda, T.; Kojo, T.; Powell, P.D.; Song, Y.; Takatsuka, T.}
\newblock {From hadrons to quarks in neutron stars: A review}.
\newblock {\em Rept. Prog. Phys.} {\bf 2018}, \emph{{81}}, 056902. [\href{http://dx.doi.org/10.1088/1361-6633/aaae14}{CrossRef}] [\href{http://www.ncbi.nlm.nih.gov/pubmed/29424363}{PubMed}]
\bibitem[Orsaria et~al.(2019)Orsaria, Malfatti, Mariani, Ranea-Sandoval,
Garc{\'{\i}}a, Spinella, Contrera, Lugones, and Weber]{Orsaria:2019pti}
Orsaria, M.G.; Malfatti, G.; Mariani, M.; Ranea-Sandoval, I.F.; Garc{\'{\i}}a,
F.; Spinella, W.M.; Contrera, G.A.; Lugones, G.; Weber, F.
\newblock Phase transitions in neutron stars and their links to gravitational
waves.
\newblock {\em J. Phys. G Nucl. Part. Phys.} {\bf 2019}, {\em 46},~073002.
\bibitem[{Dexheimer} and {Schramm}(2010)]{Dexheimer:2010ana}
{Dexheimer}, V.A.; {Schramm}, S.
\newblock {A novel approach to modeling hybrid stars}.
\newblock {\em Phys.~Rev.~C} {\bf 2010}, {\em 81},~045201. [\href{http://dx.doi.org/10.1103/PhysRevC.81.045201}{CrossRef}]
\bibitem[Kumar et~al.(2024)Kumar, Wang, Camacho, Kumar, Noronha-Hostler, and
Dexheimer]{Kumar:2024mna}
Kumar, R.; Wang, Y.; Camacho, N.C.; Kumar, A.; Noronha-Hostler, J.; Dexheimer,
V.
\newblock {Modern nuclear and astrophysical constraints of dense matter in a
redefined chiral approach}.
\newblock {\em Phys. Rev. D} {\bf 2024}, \emph{{109}}, 074008. [\href{http://dx.doi.org/10.1103/PhysRevD.109.074008}{CrossRef}]
\bibitem[Celi et~al.(2025)Celi, Mariani, Kumar, Bashkanov, Orsaria, Pastore,
Ranea-Sandoval, and Dexheimer]{celi:2025etr}
Celi, M.O.; Mariani, M.; Kumar, R.; Bashkanov, M.; Orsaria, M.G.; Pastore, A.;
Ranea-Sandoval, I.F.; Dexheimer, V.
\newblock Exploring the role of d* hexaquarks on quark deconfinement and hybrid
stars.
\newblock {\em Phys. Rev. D} {\bf 2025}, {\em 112},~023027. [\href{http://dx.doi.org/10.1103/3lyv-45jp}{CrossRef}]
\bibitem[{Logoteta} et~al.(2022){Logoteta}, {Bombaci}, and
{Perego}]{Logoteta:2022ieo}
{Logoteta}, D.; {Bombaci}, I.; {Perego}, A.
\newblock {Isoentropic equations of state of {\ensuremath{\beta}}-stable
hadronic matter with a quark phase transition}.
\newblock {\em Eur. Phys. J. A} {\bf 2022}, {\em 58},~55. [\href{http://dx.doi.org/10.1140/epja/s10050-022-00708-8}{CrossRef}]
\bibitem[Constantinou et~al.(2023)Constantinou, Zhao, Han, and
Prakash]{Constantinou:2023ffp}
Constantinou, C.; Zhao, T.; Han, S.; Prakash, M.
\newblock Framework for phase transitions between the Maxwell and Gibbs
constructions.
\newblock {\em Phys. Rev. D} {\bf 2023}, {\em 107},~074013. [\href{http://dx.doi.org/10.1103/PhysRevD.107.074013}{CrossRef}]
\bibitem[Ju et~al.(2021)Ju, Hu, and Shen]{Ju:2021hqp}
Ju, M.; Hu, J.; Shen, H.
\newblock Hadron-quark pasta phase in massive neutron stars.
\newblock {\em Astrophys. J.} {\bf 2021}, {\em 923},~250. [\href{http://dx.doi.org/10.3847/1538-4357/ac30dd}{CrossRef}]
\bibitem[Mariani and Lugones(2024)]{Mariani:2024qhp}
Mariani, M.; Lugones, G.
\newblock Quark-hadron pasta phase in neutron stars: The role of
medium-dependent surface and curvature tensions.
\newblock {\em Phys. Rev. D} {\bf 2024}, {\em 109},~063022. [\href{http://dx.doi.org/10.1103/PhysRevD.109.063022}{CrossRef}]
\bibitem[Pinto et~al.(2012)Pinto, Koch, and Randrup]{pinto:2012stq}
Pinto, M.B.; Koch, V.; Randrup, J.
\newblock Surface tension of quark matter in a geometrical approach.
\newblock {\em Phys. Rev. C---Nucl. Phys.} {\bf 2012}, {\em 86},~025203. [\href{http://dx.doi.org/10.1103/PhysRevC.86.025203}{CrossRef}]
\bibitem[Mintz et~al.(2013)Mintz, Stiele, Ramos, and
Schaffner-Bielich]{mintz:2013pda}
Mintz, B.W.; Stiele, R.; Ramos, R.O.; Schaffner-Bielich, J.
\newblock Phase diagram and surface tension in the three-flavor
Polyakov-quark-meson model.
\newblock {\em Phys. Rev. D---Part. Fields Gravit. Cosmol.} {\bf 2013}, {\em
87},~036004. [\href{http://dx.doi.org/10.1103/PhysRevD.87.036004}{CrossRef}]
\bibitem[Pereira et~al.(2018)Pereira, Flores, and Lugones]{Pereira:2017pte}
Pereira, J.P.; Flores, C.V.; Lugones, G.
\newblock {Phase transition effects on the dynamical stability of hybrid
neutron stars}.
\newblock {\em Astrophys. J.} {\bf 2018}, \emph{860},~12. [\href{http://dx.doi.org/10.3847/1538-4357/aabfbf}{CrossRef}]

\bibitem[Arbanil and Malheiro(2015)]{Arbanil:2015eas}
Arbanil, J.D.; Malheiro, M.
\newblock Equilibrium and stability of charged strange quark stars.
\newblock {\em Phys. Rev. D} {\bf 2015}, {\em 92},~084009. [\href{http://dx.doi.org/10.1103/PhysRevD.92.084009}{CrossRef}]
\bibitem[Mohanty et~al.(2024)Mohanty, Ghosh, and Kumar]{mohanty:2024uan}
Mohanty, S.R.; Ghosh, S.; Kumar, B.
\newblock Unstable anisotropic neutron stars: Probing the limits of
gravitational collapse.
\newblock {\em Phys. Rev. D} {\bf 2024}, {\em 109},~123039. [\href{http://dx.doi.org/10.1103/PhysRevD.109.123039}{CrossRef}]
\bibitem[Kain(2020)]{kain:2020roa}
Kain, B.
\newblock Radial oscillations and stability of multiple-fluid compact stars.
\newblock {\em Phys. Rev. D} {\bf 2020}, {\em 102},~023001. [\href{http://dx.doi.org/10.1103/PhysRevD.102.023001}{CrossRef}]
\bibitem[Caballero et~al.(2024)Caballero, Ripley, and Yunes]{caballero:2024rms}
Caballero, D.A.; Ripley, J.; Yunes, N.
\newblock Radial mode stability of two-fluid neutron stars.
\newblock {\em Phys. Rev. D} {\bf 2024}, {\em 110},~103038. [\href{http://dx.doi.org/10.1103/PhysRevD.110.103038}{CrossRef}]
\bibitem[Pereira et~al.(2021)Pereira, Bejger, Tonetto, Lugones, Haensel,
Zdunik, and Sieniawska]{pereira:2021peq}
Pereira, J.P.; Bejger, M.; Tonetto, L.; Lugones, G.; Haensel, P.; Zdunik, J.L.;
Sieniawska, M.
\newblock Probing elastic quark phases in hybrid stars with radius
measurements.
\newblock {\em Astrophys. J.} {\bf 2021}, {\em 910},~145. [\href{http://dx.doi.org/10.3847/1538-4357/abe633}{CrossRef}]
\bibitem[Canullan-Pascual et~al.(2025)Canullan-Pascual, Lugones, Orsaria, and
Ranea-Sandoval]{canullan:2025nss}
Canullan-Pascual, M.O.; Lugones, G.; Orsaria, M.G.; Ranea-Sandoval, I.F.
\newblock Neutron star stability beyond the mass peak: Assessing the role of
out-of-equilibrium perturbations.
\newblock {\em Astrophys. J.} {\bf 2025}, {\em 989},~135. [\href{http://dx.doi.org/10.3847/1538-4357/adf107}{CrossRef}]
\bibitem[Drago et~al.(2014)Drago, Lavagno, and Pagliara]{Drago:2014cvc}
Drago, A.; Lavagno, A.; Pagliara, G.
\newblock Can very compact and very massive neutron stars both exist?
\newblock {\em Phys. Rev. D} {\bf 2014}, {\em 89},~043014. [\href{http://dx.doi.org/10.1103/PhysRevD.89.043014}{CrossRef}]
\bibitem[{Di Clemente} et~al.(2024){Di Clemente}, {Drago}, and
{Pagliara}]{DiClemente:2024itc}
{Di Clemente}, F.; {Drago}, A.; {Pagliara}, G.
\newblock {Is the Compact Object Associated with HESS J1731-347 a Strange Quark
Star? A Possible Astrophysical Scenario for Its Formation}.
\newblock {\em Astrophys. J. Lett.} {\bf 2024}, {\em 967},~159. [\href{http://dx.doi.org/10.3847/1538-4357/ad445b}{CrossRef}]
\bibitem[{Drago} et~al.(2016){Drago}, {Lavagno}, {Pagliara}, and
{Pigato}]{Drago:2016tso1}
{Drago}, A.; {Lavagno}, A.; {Pagliara}, G.; {Pigato}, D.
\newblock {The scenario of two families of compact stars. Part 1. Equations of
state, mass-radius relations and binary systems}.
\newblock {\em Eur. Phys. J. A} {\bf 2016}, {\em 52},~40. [\href{http://dx.doi.org/10.1140/epja/i2016-16040-3}{CrossRef}]
\bibitem[Drago and Pagliara(2016)]{Drago:2016tso2}
Drago, A.; Pagliara, G.
\newblock The scenario of two families of compact stars: Part 2: Transition
from hadronic to quark matter and explosive phenomena.
\newblock {\em Eur. Phys. J. A} {\bf 2016}, {\em 52},~41. [\href{http://dx.doi.org/10.1140/epja/i2016-16041-2}{CrossRef}]
\bibitem[Abgaryan et~al.(2018)Abgaryan, Alvarez-Castillo, Ayriyan, Blaschke,
and Grigorian]{abgaryan2018two}
Abgaryan, V.; Alvarez-Castillo, D.; Ayriyan, A.; Blaschke, D.; Grigorian, H.
\newblock Two novel approaches to the Hadron-Quark mixed phase in compact
stars.
\newblock {\em Universe} {\bf 2018}, {\em 4},~94. [\href{http://dx.doi.org/10.3390/universe4090094}{CrossRef}]
\bibitem[Ranea-Sandoval et~al.(2019)Ranea-Sandoval, Orsaria, Malfatti, Curin,
Mariani, Contrera, and Guilera]{ranea:2019eoh}
Ranea-Sandoval, I.F.; Orsaria, M.G.; Malfatti, G.; Curin, D.; Mariani, M.;
Contrera, G.A.; Guilera, O.M.
\newblock Effects of hadron-quark phase transitions in hybrid stars within the
NJL model.
\newblock {\em Symmetry} {\bf 2019}, {\em 11},~425. [\href{http://dx.doi.org/10.3390/sym11030425}{CrossRef}]
\bibitem[Carlomagno et~al.(2024)Carlomagno, Contrera, Grunfeld, and
Blaschke]{carlomagno:2024tts}
Carlomagno, J.P.; Contrera, G.A.; Grunfeld, A.G.; Blaschke, D.
\newblock Thermal twin stars within a hybrid equation of state based on a
nonlocal chiral quark model compatible with modern astrophysical
observations.
\newblock {\em Phys. Rev. D} {\bf 2024}, {\em 109},~043050. [\href{http://dx.doi.org/10.1103/PhysRevD.109.043050}{CrossRef}]
\bibitem[Alvarez-Castillo et~al.(2019)Alvarez-Castillo, Blaschke, Grunfeld, and
Pagura]{alvarez:2019tfo}
Alvarez-Castillo, D.; Blaschke, D.; Grunfeld, A.G.; Pagura, V.
\newblock Third family of compact stars within a nonlocal chiral quark model
equation of state.
\newblock {\em Phys. Rev. D} {\bf 2019}, {\em 99},~063010. [\href{http://dx.doi.org/10.1103/PhysRevD.99.063010}{CrossRef}]
\bibitem[Pal and Chaudhuri(2025)]{Pal:2025cah}
Pal, S.; Chaudhuri, G.
\newblock Can a Hybrid Star with Constant Sound Speed Parameterization Explain
the New NICER Mass–Radius Measurements?
\newblock {\em Astrophys. J.} {\bf 2025}, {\em 991},~158. [\href{http://dx.doi.org/10.3847/1538-4357/adf6a7}{CrossRef}]
\bibitem[Alford and Sedrakian(2017)]{alford:2017csw}
Alford, M.; Sedrakian, A.
\newblock Compact Stars with Sequential QCD Phase Transitions.
\newblock {\em Phys. Rev. Lett.} {\bf 2017}, {\em 119},~161104. [\href{http://dx.doi.org/10.1103/PhysRevLett.119.161104}{CrossRef}]
\bibitem[Rodr{\'\i}guez et~al.(2021)Rodr{\'\i}guez, Ranea-Sandoval, Mariani,
Orsaria, Malfatti, and Guilera]{rodriguez:2021hsw}
Rodr{\'\i}guez, M.C.; Ranea-Sandoval, I.F.; Mariani, M.; Orsaria, M.G.;
Malfatti, G.; Guilera, O.M.
\newblock Hybrid stars with sequential phase transitions: The emergence of the
g2 mode.
\newblock {\em J. Cosmol. Astropart. Phys.} {\bf 2021}, {\em 2021},~009. [\href{http://dx.doi.org/10.1088/1475-7516/2021/02/009}{CrossRef}]
\bibitem[Li et~al.(2023)Li, Sedrakian, and Alford]{li:2023rhs}
Li, J.J.; Sedrakian, A.; Alford, M.
\newblock Relativistic hybrid stars with sequential first-order phase
transitions in light of multimessenger constraints.
\newblock {\em Astrophys. J.} {\bf 2023}, {\em 944},~206. [\href{http://dx.doi.org/10.3847/1538-4357/acb688}{CrossRef}]
\bibitem[Gon{\c{c}}alves and Lazzari(2022)]{Goncalves:2022ios}
Gon{\c{c}}alves, V.P.; Lazzari, L.
\newblock Impact of slow conversions on hybrid stars with sequential QCD phase
transitions.
\newblock {\em Eur. Phys. J. C} {\bf 2022}, {\em 82},~288. [\href{http://dx.doi.org/10.1140/epjc/s10052-022-10273-5}{CrossRef}]
\bibitem[Lugones et~al.(2023)Lugones, Mariani, and
Ranea-Sandoval]{Lugones:2023ama}
Lugones, G.; Mariani, M.; Ranea-Sandoval, I.F.
\newblock A model-agnostic analysis of hybrid stars with reactive interfaces.
\newblock {\em J. Cosmol. Astropart. Phys.} {\bf 2023}, {\em 2023},~028. [\href{http://dx.doi.org/10.1088/1475-7516/2023/03/028}{CrossRef}]
\bibitem[{Rau} and {Salaben}(2023)]{Rau:2023neo}
{Rau}, P.B.; {Salaben}, G.G.
\newblock {Nonequilibrium effects on stability of hybrid stars with first-order
phase transitions}.
\newblock {\em Phys. Rev. D} {\bf 2023}, {\em 108},~103035. [\href{http://dx.doi.org/10.1103/physrevd.108.103035}{CrossRef}]
\bibitem[Rau and Sedrakian(2023)]{Rau:2023tfo}
Rau, P.B.; Sedrakian, A.
\newblock Two first-order phase transitions in hybrid compact stars:
Higher-order multiplet stars, reaction modes, and intermediate conversion
speeds.
\newblock {\em Phys. Rev. D} {\bf 2023}, {\em 107},~103042. [\href{http://dx.doi.org/10.1103/PhysRevD.107.103042}{CrossRef}]
\bibitem[Mariani et~al.(2024)Mariani, Ranea-Sandoval, Lugones, and
Orsaria]{Mariani:2024cas}
Mariani, M.; Ranea-Sandoval, I.F.; Lugones, G.; Orsaria, M.G.
\newblock Could a slow stable hybrid star explain the central compact object in
HESS J1731-347?
\newblock {\em Phys. Rev. D} {\bf 2024}, {\em 110},~043026. [\href{http://dx.doi.org/10.1103/PhysRevD.110.043026}{CrossRef}]
\bibitem[Laskos-Patkos and Moustakidis(2025)]{Laskos:2025xja}
Laskos-Patkos, P.; Moustakidis, C.C.
\newblock XTE J1814-338: A potential hybrid star candidate.
\newblock {\em Phys. Rev. D} {\bf 2025}, {\em 111},~063058. [\href{http://dx.doi.org/10.1103/PhysRevD.111.063058}{CrossRef}]
\bibitem[{Zhang} et~al.(2025){Zhang}, {Pretel}, and {Xu}]{Zhang:2025sss}
{Zhang}, C.; {Pretel}, J.M.Z.; {Xu}, R.
\newblock {Slow Stable Self-bound Hybrid Star Can Relieve All Tensions}.
\newblock {\em arXiv} {\bf 2025}, arXiv:2507.01371. [\href{http://dx.doi.org/10.48550/arXiv.2507.01371}{CrossRef}]
\bibitem[{Yang} et~al.(2025){Yang}, {Zeng}, {Yan}, {Yuan}, {Zhang}, and
{Zhou}]{Yang:2025hqs}
{Yang}, Z.; {Zeng}, T.; {Yan}, Y.; {Yuan}, W.L.; {Zhang}, C.; {Zhou}, E.
\newblock {Hybrid Quark Stars with Quark-Quark Phase Transitions}.
\newblock {\em arXiv} {\bf 2025}, arXiv:2507.00776. [\href{http://dx.doi.org/10.48550/arXiv.2507.00776}{CrossRef}]
\bibitem[{Sagun} et~al.(2023){Sagun}, {Giangrandi}, {Dietrich}, {Ivanytskyi},
{Negreiros}, and {Provid{\^e}ncia}]{Sagun:2023wit}
{Sagun}, V.; {Giangrandi}, E.; {Dietrich}, T.; {Ivanytskyi}, O.; {Negreiros},
R.; {Provid{\^e}ncia}, C.
\newblock {What Is the Nature of the HESS J1731-347 Compact Object?}
\newblock {\em Astrophys. J. Lett.} {\bf 2023}, {\em 958},~49. [\href{http://dx.doi.org/10.3847/1538-4357/acfc9e}{CrossRef}]

\bibitem[Alford et~al.(2001)Alford, Bowers, and Rajagopal]{alford:2001ccc}
Alford, M.; Bowers, J.A.; Rajagopal, K.
\newblock Crystalline color superconductivity.
\newblock {\em Phys. Rev. D} {\bf 2001}, {\em 63},~074016. [\href{http://dx.doi.org/10.1103/PhysRevD.63.074016}{CrossRef}]
\bibitem[Alford(2001)]{alford:2001csq}
Alford, M.
\newblock Color-superconducting quark matter.
\newblock {\em Annu. Rev. Nucl. Part. Sci.} {\bf 2001}, {\em 51},~131--160. [\href{http://dx.doi.org/10.1146/annurev.nucl.51.101701.132449}{CrossRef}]
\bibitem[Ruester et~al.(2005)Ruester, Werth, Buballa, Shovkovy, and
Rischke]{ruester:2005pdo}
Ruester, S.B.; Werth, V.; Buballa, M.; Shovkovy, I.A.; Rischke, D.H.
\newblock Phase diagram of neutral quark matter: Self-consistent treatment of
quark masses.
\newblock {\em Phys. Rev. D—Part. Fields Gravit. Cosmol.} {\bf 2005}, {\em
72},~034004. [\href{http://dx.doi.org/10.1103/PhysRevD.72.034004}{CrossRef}]
\bibitem[{Miralda-Escude} et~al.(1990){Miralda-Escude}, {Haensel}, and
{Paczynski}]{Miralda:1990tso}
{Miralda-Escude}, J.; {Haensel}, P.; {Paczynski}, B.
\newblock {Thermal Structure of Accreting Neutron Stars and Strange Stars}.
\newblock {\em Astrophys. J. Lett.} {\bf 1990}, {\em 362},~572. [\href{http://dx.doi.org/10.1086/169295}{CrossRef}]
\bibitem[{O'Boyle} et~al.(2020){O'Boyle}, {Markakis}, {Stergioulas}, and
{Read}]{OBoyle:2020peo}
{O'Boyle}, M.F.; {Markakis}, C.; {Stergioulas}, N.; {Read}, J.S.
\newblock {Parametrized equation of state for neutron star matter with
continuous sound speed}.
\newblock {\em Phys. Rev. D} {\bf 2020}, {\em 102},~083027. [\href{http://dx.doi.org/10.1103/physrevd.102.083027}{CrossRef}]
\bibitem[{Baym} et~al.(1971){Baym}, {Pethick}, and {Sutherland}]{Baym:1971tgs}
{Baym}, G.; {Pethick}, C.; {Sutherland}, P.
\newblock {The Ground State of Matter at High Densities: Equation of State and
Stellar Models}.
\newblock {\em Astrophys. J. Lett.} {\bf 1971}, {\em 170},~299. [\href{http://dx.doi.org/10.1086/151216}{CrossRef}]
\bibitem[Baym et~al.(1971)Baym, Bethe, and Pethick]{Baym:1971nsm}
Baym, G.; Bethe, H.A.; Pethick, C.J.
\newblock Neutron star matter.
\newblock {\em Nucl. Phys. A} {\bf 1971}, {\em 175},~225--271. [\href{http://dx.doi.org/10.1016/0375-9474(71)90281-8}{CrossRef}]
\bibitem[Johnson et~al.(1975)]{johnson:1975tmb}
Johnson, K.
\newblock The MIT bag model.
\newblock {\em Acta Phys. Pol. B} {\bf 1975}, {\em 6},~8.
\bibitem[Buballa(2005)]{buballa:2005nma}
Buballa, M.
\newblock NJL-model analysis of dense quark matter.
\newblock {\em Phys. Rep.} {\bf 2005}, {\em 407},~205--376. [\href{http://dx.doi.org/10.1016/j.physrep.2004.11.004}{CrossRef}]
\bibitem[Orsaria et~al.(2014)Orsaria, Rodrigues, Weber, and
Contrera]{orsaria:2014qdi}
Orsaria, M.; Rodrigues, H.; Weber, F.; Contrera, G.
\newblock Quark deconfinement in high-mass neutron stars.
\newblock {\em Phys. Rev. C} {\bf 2014}, {\em 89},~015806. [\href{http://dx.doi.org/10.1103/PhysRevC.89.015806}{CrossRef}]
\bibitem[Malfatti et~al.(2020)Malfatti, Orsaria, Ranea-Sandoval, Contrera, and
Weber]{malfatti:2020dba}
Malfatti, G.; Orsaria, M.G.; Ranea-Sandoval, I.F.; Contrera, G.A.; Weber, F.
\newblock Delta baryons and diquark formation in the cores of neutron stars.
\newblock {\em Phys. Rev. D} {\bf 2020}, {\em 102},~063008. [\href{http://dx.doi.org/10.1103/PhysRevD.102.063008}{CrossRef}]
\bibitem[Plumari et~al.(2013)Plumari, Burgio, Greco, and
Zappala]{plumari:2013qmi}
Plumari, S.; Burgio, G.; Greco, V.; Zappala, D.
\newblock Quark matter in neutron stars within the field correlator method.
\newblock {\em Phys. Rev. D—Part. Fields Gravit. Cosmol.} {\bf 2013}, {\em
88},~083005. [\href{http://dx.doi.org/10.1103/PhysRevD.88.083005}{CrossRef}]
\bibitem[{Ranea-Sandoval} et~al.(2016){Ranea-Sandoval}, {Han}, {Orsaria},
{Contrera}, {Weber}, and {Alford}]{ranea:2016css}
{Ranea-Sandoval}, I.F.; {Han}, S.; {Orsaria}, M.G.; {Contrera}, G.A.; {Weber},
F.; {Alford}, M.G.
\newblock {Constant-sound-speed parametrization for Nambu-Jona-Lasinio models
of quark matter in hybrid stars}.
\newblock {\em Phys. Rev. C} {\bf 2016}, {\em 93},~045812. [\href{http://dx.doi.org/10.1103/PhysRevC.93.045812}{CrossRef}]
\bibitem[{Mariani} et~al.(2017){Mariani}, {Orsaria}, and
{Vucetich}]{mariani:2017ceh}
{Mariani}, M.; {Orsaria}, M.; {Vucetich}, H.
\newblock {Constant entropy hybrid stars: A first approximation of cooling
evolution}.
\newblock {\em Astron. Astrophys.} {\bf 2017}, {\em 601},~A21. [\href{http://dx.doi.org/10.1051/0004-6361/201629315}{CrossRef}]
\bibitem[Celi et~al.(2025)Celi, Mariani, Orsaria, Ranea-Sandoval, and
Lugones]{celi:2025tau-arxiv}
{Celi, M.O.; Mariani, M.; Orsaria, M.G.; Ranea-Sandoval, I.F.; Lugones, G.}
\newblock Towards a unified hadron-quark equation of state for neutron stars
within the relativistic mean-field model.
\newblock {\em {Phys. Rev. D}} {\bf 2025}, \emph{112}, 123001. [\href{http://dx.doi.org/10.1103/ynml-q8zm}{CrossRef}]
\bibitem[Wang et~al.(2019)Wang, Shi, and Zong]{wang:2019nqs}
Wang, Q.; Shi, C.; Zong, H.S.
\newblock Nonstrange quark stars from an NJL model with proper-time
regularization.
\newblock {\em Phys. Rev. D} {\bf 2019}, {\em 100},~123003. [\href{http://dx.doi.org/10.1103/PhysRevD.100.123003}{CrossRef}]
\bibitem[Li et~al.(2022)Li, Zuo, Zhao, Mu, and Huang]{li:2022son}
Li, C.M.; Zuo, S.Y.; Zhao, Y.P.; Mu, H.J.; Huang, Y.F.
\newblock Study of nonstrange quark stars within a modified NJL model.
\newblock {\em Phys. Rev. D} {\bf 2022}, {\em 106},~116009. [\href{http://dx.doi.org/10.1103/PhysRevD.106.116009}{CrossRef}]
\bibitem[{Lopes} et~al.(2021){Lopes}, {Biesdorf}, and {Menezes}]{Lopes:2021mmb}
{Lopes}, L.L.; {Biesdorf}, C.; {Menezes}, D.P.
\newblock {Modified MIT bag Models{\textemdash}Part I: Thermodynamic
consistency, stability windows and symmetry group}.
\newblock {\em Phys. Scr.} {\bf 2021}, {\em 96},~065303. [\href{http://dx.doi.org/10.1088/1402-4896/abef34}{CrossRef}]
\bibitem[{Ju} et~al.(2025){Ju}, {Chu}, {Wu}, and {Liu}]{ju:2025hji}
{Ju}, M.; {Chu}, P.; {Wu}, X.; {Liu}, H.
\newblock {Hess J1731-347 is likely a quark star based on the density-dependent
vMIT bag model}.
\newblock {\em Eur. Phys. J. C} {\bf 2025}, {\em 85},~40. [\href{http://dx.doi.org/10.1140/epjc/s10052-025-13788-9}{CrossRef}]
\bibitem[Alford et~al.(2005)Alford, Braby, Paris, and Reddy]{alford:2005hst}
Alford, M.; Braby, M.; Paris, M.; Reddy, S.
\newblock Hybrid stars that masquerade as neutron stars.
\newblock {\em Astrophys. J.} {\bf 2005}, {\em 629},~969. [\href{http://dx.doi.org/10.1086/430902}{CrossRef}]
\bibitem[Lugones and Grunfeld(2024)]{lugones:2024eve}
Lugones, G.; Grunfeld, A.G.
\newblock Excluded volume effects in the quark-mass density-dependent model:
Implications for the equation of state and compact star structure.
\newblock {\em Phys. Rev. D} {\bf 2024}, {\em 109},~063025. [\href{http://dx.doi.org/10.1103/PhysRevD.109.063025}{CrossRef}]
\bibitem[Beni{\'c}(2014)]{benic:2014hhs}
Beni{\'c}, S.
\newblock Heavy hybrid stars from multi-quark interactions.
\newblock {\em Eur. Phys. J. A} {\bf 2014}, {\em 50},~111. [\href{http://dx.doi.org/10.1140/epja/i2014-14111-1}{CrossRef}]
\bibitem[{Bedaque} and {Steiner}(2015)]{Bedaque:2015svb}
{Bedaque}, P.; {Steiner}, A.W.
\newblock {Sound Velocity Bound and Neutron Stars}.
\newblock {\em Phys. Rev. Lett.} {\bf 2015}, {\em 114},~031103. [\href{http://dx.doi.org/10.1103/PhysRevLett.114.031103}{CrossRef}]
\bibitem[Otto et~al.(2020)Otto, Oertel, and Schaefer]{otto:2020haq}
Otto, K.; Oertel, M.; Schaefer, B.J.
\newblock Hybrid and quark star matter based on a nonperturbative equation of
state.
\newblock {\em Phys. Rev. D} {\bf 2020}, {\em 101},~103021. [\href{http://dx.doi.org/10.1103/PhysRevD.101.103021}{CrossRef}]
\bibitem[Rather et~al.(2020)Rather, Kumar, Das, Imran, Usmani, and
Patra]{rather:2020cbc}
Rather, I.A.; Kumar, A.; Das, H.C.; Imran, M.; Usmani, A.; Patra, S.K.
\newblock Constraining bag constant for hybrid neutron stars.
\newblock {\em Int. J. Mod. Phys. E} {\bf 2020}, {\em 29},~2050044. [\href{http://dx.doi.org/10.1142/S0218301320500445}{CrossRef}]
\bibitem[{Mitra} et~al.(2022){Mitra}, {Sahoo}, {Mishra}, and
{Panda}]{mitra:2022hsi}
{Mitra}, G.; {Sahoo}, H.S.; {Mishra}, R.; {Panda}, P.K.
\newblock {Hybrid stars in a relativistic quark model}.
\newblock {\em Phys.~Rev.~C} {\bf 2022}, {\em 105},~045802. [\href{http://dx.doi.org/10.1103/PhysRevC.105.045802}{CrossRef}]
\bibitem[Karimi and Moshfegh(2023)]{karimi:2023hsw}
Karimi, R.; Moshfegh, H.
\newblock Hybrid stars within the framework of the Sigma-Omega-Rho model
combined with the MIT and NJL models.
\newblock {\em Nucl. Phys. A} {\bf 2023}, {\em 1037},~122684. [\href{http://dx.doi.org/10.1016/j.nuclphysa.2023.122684}{CrossRef}]
\bibitem[Song(2025)]{song:2025csa}
Song, X.Y.
\newblock Compact stars as hideouts for color-spin-locked quark matter:
Implications for powering high-energy electromagnetic emissions.
\newblock {\em Phys. Rev. D} {\bf 2025}, {\em 111},~063018. [\href{http://dx.doi.org/10.1103/PhysRevD.111.063018}{CrossRef}]
\bibitem[Alford et~al.(2013)Alford, Han, and Prakash]{alford:2013gcf}
Alford, M.G.; Han, S.; Prakash, M.
\newblock Generic conditions for stable hybrid stars.
\newblock {\em Phys. Rev. D—Part. Fields Gravit. Cosmol.} {\bf 2013}, {\em
88},~083013.

\bibitem[Ivanytskyi and Blaschke(2022)]{ivanytskyi:2022rtc}
Ivanytskyi, O.; Blaschke, D.B.
\newblock Recovering the Conformal Limit of Color Superconducting Quark Matter
within a Confining Density Functional Approach.
\newblock {\em Particles} {\bf 2022}, {\em 5},~514--534. [\href{http://dx.doi.org/10.3390/particles5040038}{CrossRef}]
\bibitem[Lugones and Grunfeld(2024)]{lugones:2024sqs}
Lugones, G.; Grunfeld, A.G.
\newblock Strange Quark Stars: The Role of Excluded Volume Effects.
\newblock {\em Universe} {\bf 2024}, {\em 10}, {233}. [\href{http://dx.doi.org/10.3390/universe10060233}{CrossRef}]
\bibitem[{Traversi, Silvia} et~al.(2022){Traversi, Silvia}, {Char, Prasanta},
{Pagliara, Giuseppe}, and {Drago, Alessandro}]{Traversi:2022sos}
Traversi, S.; Char, P.; Pagliara, G.; Drago, A.
\newblock Speed of sound in dense matter and two families of compact stars.
\newblock {\em Astron. Astrophys.} {\bf 2022}, {\em 660},~A62. [\href{http://dx.doi.org/10.1051/0004-6361/202141544}{CrossRef}]
\bibitem[Gorda et~al.(2023)Gorda, Komoltsev, Kurkela, and
Mazeliauskas]{gorda:2023buq}
Gorda, T.; Komoltsev, O.; Kurkela, A.; Mazeliauskas, A.
\newblock Bayesian uncertainty quantification of perturbative QCD input to the
neutron-star equation of state.
\newblock {\em J. High Energy Phys.} {\bf 2023}, {\em 2023},~{2}. [\href{http://dx.doi.org/10.1007/JHEP06(2023)002}{CrossRef}]
\bibitem[Komoltsev et~al.(2024)Komoltsev, Somasundaram, Gorda, Kurkela,
Margueron, and Tews]{komoltsev:2024eos}
Komoltsev, O.; Somasundaram, R.; Gorda, T.; Kurkela, A.; Margueron, J.; Tews,
I.
\newblock Equation of state at neutron-star densities and beyond from
perturbative QCD.
\newblock {\em Phys. Rev. D} {\bf 2024}, {\em 109},~094030. [\href{http://dx.doi.org/10.1103/PhysRevD.109.094030}{CrossRef}]
\bibitem[Hippert et~al.(2025)Hippert, Noronha, and Romatschke]{hippert:2025ubo}
Hippert, M.; Noronha, J.; Romatschke, P.
\newblock Upper bound on the speed of sound in nuclear matter from transport.
\newblock {\em Phys. Lett. B} {\bf 2025}, {\em 860},~139184. [\href{http://dx.doi.org/10.1016/j.physletb.2024.139184}{CrossRef}]
\bibitem[Kojo(2021)]{kojo:2021qeo}
Kojo, T.
\newblock QCD equations of state and speed of sound in neutron stars.
\newblock {\em AAPPS Bull.} {\bf 2021}, {\em 31},~11. [\href{http://dx.doi.org/10.1007/s43673-021-00011-6}{CrossRef}]
\bibitem[{Lugones} and {Grunfeld}(2021)]{Lugones:2021pci}
{Lugones}, G.; {Grunfeld}, A.G.
\newblock {Phase Conversions in Neutron Stars: Implications for Stellar
Stability and Gravitational Wave Astrophysics}.
\newblock {\em Universe} {\bf 2021}, {\em 7},~493. [\href{http://dx.doi.org/10.3390/universe7120493}{CrossRef}]
\bibitem[{Ranea-Sandoval} et~al.(2023){Ranea-Sandoval}, {Mariani}, {Celi},
{Rodr{\'\i}guez}, and {Tonetto}]{Ranea:2023auq}
{Ranea-Sandoval}, I.F.; {Mariani}, M.; {Celi}, M.O.; {Rodr{\'\i}guez}, M.C.;
{Tonetto}, L.
\newblock {Asteroseismology using quadrupolar f -modes revisited: Breaking of
universal relationships in the slow hadron-quark conversion scenario}.
\newblock {\em Phys. Rev. D} {\bf 2023}, {\em 107},~123028. [\href{http://dx.doi.org/10.1103/PhysRevD.107.123028}{CrossRef}]
\bibitem[Ranea-Sandoval et~al.(2022)Ranea-Sandoval, Guilera, Mariani, and
Lugones]{ranea:2022bou}
Ranea-Sandoval, I.F.; Guilera, O.M.; Mariani, M.; Lugones, G.
\newblock Breaking of universal relationships of axial w I modes in hybrid
stars: Rapid and slow hadron-quark conversion scenarios.
\newblock {\em Phys. Rev. D} {\bf 2022}, {\em 106},~043025. [\href{http://dx.doi.org/10.1103/PhysRevD.106.043025}{CrossRef}]
\bibitem[{Mariani} et~al.(2024){Mariani}, {Albertus}, {Alessandroni},
{Orsaria}, {P{\'e}rez-Garc{\'\i}a}, and {Ranea-Sandoval}]{Mariani:2024csi}
{Mariani}, M.; {Albertus}, C.; {Alessandroni}, M.d.R.; {Orsaria}, M.G.;
{P{\'e}rez-Garc{\'\i}a}, M.{\'A}.; {Ranea-Sandoval}, I.F.
\newblock {Constraining self-interacting fermionic dark matter in admixed
neutron stars using multimessenger astronomy}.
\newblock {\em MNRAS} {\bf 2024}, {\em 527},~6795--6806. [\href{http://dx.doi.org/10.1093/mnras/stad3658}{CrossRef}]
\bibitem[{Menezes} and {Melrose}(2005)]{Menezes:2005sse}
{Menezes}, D.P.; {Melrose}, D.B.
\newblock {Strange Star Equations of State Revisited}.
\newblock {\em PASA} {\bf 2005}, {\em 22},~292--297. [\href{http://dx.doi.org/10.1071/AS05022}{CrossRef}]
\bibitem[{Christian} et~al.(2018){Christian}, {Zacchi}, and
{Schaffner-Bielich}]{Christian:2018cot}
{Christian}, J.E.; {Zacchi}, A.; {Schaffner-Bielich}, J.
\newblock {Classifications of twin star solutions for a constant speed of sound
parameterized equation of state}.
\newblock {\em Eur. Phys. J. A} {\bf 2018}, {\em 54},~28. [\href{http://dx.doi.org/10.1140/epja/i2018-12472-y}{CrossRef}]
\bibitem[Ranea-Sandoval et~al.(2017)Ranea-Sandoval, Orsaria, Han, Weber, and
Spinella]{ranea:2017csi}
Ranea-Sandoval, I.F.; Orsaria, M.G.; Han, S.; Weber, F.; Spinella, W.M.
\newblock Color superconductivity in compact stellar hybrid configurations.
\newblock {\em Phys. Rev. C} {\bf 2017}, {\em 96},~065807. [\href{http://dx.doi.org/10.1103/PhysRevC.96.065807}{CrossRef}]
\bibitem[Tonetto and Lugones(2020)]{tonetto:2020dgm}
Tonetto, L.; Lugones, G.
\newblock Discontinuity gravity modes in hybrid stars: Assessing the role of
rapid and slow phase conversions.
\newblock {\em Phys. Rev. D} {\bf 2020}, {\em 101},~123029. [\href{http://dx.doi.org/10.1103/physrevd.101.123029}{CrossRef}]
\bibitem[Rather et~al.(2024)Rather, Marquez, Backes, Panotopoulos, and
Lopes]{rather:2024roo}
Rather, I.A.; Marquez, K.D.; Backes, B.C.; Panotopoulos, G.; Lopes, I.
\newblock Radial oscillations of hybrid stars and neutron stars including delta
baryons: The effect of a slow quark phase transition.
\newblock {\em J. Cosmol. Astropart. Phys.} {\bf 2024}, {\em 2024},~130. [\href{http://dx.doi.org/10.1088/1475-7516/2024/05/130}{CrossRef}]
\bibitem[Mariani et~al.(2019)Mariani, Orsaria, Ranea-Sandoval, and
Lugones]{mariani:2019mhs}
Mariani, M.; Orsaria, M.G.; Ranea-Sandoval, I.F.; Lugones, G.
\newblock Magnetized hybrid stars: Effects of slow and rapid phase transitions
at the quark--hadron interface.
\newblock {\em Mon. Not. R. Astron. Soc.} {\bf 2019}, {\em 489},~4261--4277. [\href{http://dx.doi.org/10.1093/mnras/stz2392}{CrossRef}]
\bibitem[Lenzi et~al.(2023)Lenzi, Lugones, and Vasquez]{lenzi:2023hsw}
Lenzi, C.; Lugones, G.; Vasquez, C.
\newblock Hybrid stars with reactive interfaces: Analysis within the
Nambu--Jona-Lasinio model.
\newblock {\em Phys. Rev. D} {\bf 2023}, {\em 107},~083025. [\href{http://dx.doi.org/10.1103/PhysRevD.107.083025}{CrossRef}]
\bibitem[Shibata et~al.(2019)Shibata, Zhou, Kiuchi, and
Fujibayashi]{Shibata:2019ctb}
Shibata, M.; Zhou, E.; Kiuchi, K.; Fujibayashi, S.
\newblock {Constraint on the maximum mass of neutron stars using GW170817
event}.
\newblock {\em Phys. Rev. D} {\bf 2019}, \emph{{100}}, 023015. [\href{http://dx.doi.org/10.1103/physrevd.100.023015}{CrossRef}]
\bibitem[Fortin et~al.(2016)Fortin, Provid{\^e}ncia, Raduta, Gulminelli,
Zdunik, Haensel, and Bejger]{Fortin:2016nsr}
Fortin, M.; Provid{\^e}ncia, C.; Raduta, A.R.; Gulminelli, F.; Zdunik, J.;
Haensel, P.; Bejger, M.
\newblock Neutron star radii and crusts: Uncertainties and unified equations of
state.
\newblock {\em Phys. Rev. C} {\bf 2016}, {\em 94},~035804. [\href{http://dx.doi.org/10.1103/physrevc.94.035804}{CrossRef}]
\bibitem[Canull{\'a}n-Pascual et~al.(2025)Canull{\'a}n-Pascual, Mariani,
Ranea-Sandoval, Orsaria, and Weber]{Canullan:2025ccc}
Canull{\'a}n-Pascual, M.O.; Mariani, M.; Ranea-Sandoval, I.F.; Orsaria, M.G.;
Weber, F.
\newblock Consistent Crust-Core Interpolation and Its Effect on Non-radial
Neutron Star Oscillations.
\newblock {\em Astron. Nachrichten} {\bf 2025}, {\em 346},~e20240150. [\href{http://dx.doi.org/10.1002/asna.20240150}{CrossRef}]
\bibitem[{Shirke} et~al.(2025){Shirke}, {Maiti}, and
{Chatterjee}]{Shirke:2025pan}
{Shirke}, S.; {Maiti}, R.; {Chatterjee}, D.
\newblock {PSR J0614-3329: A NICER case for Strange Quark Stars}.
\newblock {\em arXiv} {\bf 2025}, arXiv:2508.02652. [\href{http://dx.doi.org/10.48550/arXiv.2508.02652}{CrossRef}]
\bibitem[Sotani et~al.(2001)Sotani, Tominaga, and Maeda]{sotani:2001ddo}
Sotani, H.; Tominaga, K.; Maeda, K.i.
\newblock Density discontinuity of a neutron star and gravitational waves.
\newblock {\em Phys. Rev. D} {\bf 2001}, {\em 65},~024010. [\href{http://dx.doi.org/10.1103/physrevd.65.024010}{CrossRef}]
\bibitem[Miniutti et~al.(2003)Miniutti, Pons, Berti, Gualtieri, and
Ferrari]{miniutti:2003nro}
Miniutti, G.; Pons, J.; Berti, E.; Gualtieri, L.; Ferrari, V.
\newblock Non-radial oscillation modes as a probe of density discontinuities in
neutron stars.
\newblock {\em Mon. Not. R. Astron. Soc.} {\bf 2003}, {\em 338},~389--400. [\href{http://dx.doi.org/10.1046/j.1365-8711.2003.06057.x}{CrossRef}]
\bibitem[Ranea-Sandoval et~al.(2018)Ranea-Sandoval, Guilera, Mariani, and
Orsaria]{ranea:2018omo}
Ranea-Sandoval, I.F.; Guilera, O.M.; Mariani, M.; Orsaria, M.G.
\newblock Oscillation modes of hybrid stars within the relativistic Cowling
approximation.
\newblock {\em J. Cosmol. Astropart. Phys.} {\bf 2018}, {\em 2018},~031. [\href{http://dx.doi.org/10.1088/1475-7516/2018/12/031}{CrossRef}]
\bibitem[Ranea-Sandoval et~al.(2023)Ranea-Sandoval, Mariani, Lugones, and
Guilera]{ranea:2023cmr}
Ranea-Sandoval, I.F.; Mariani, M.; Lugones, G.; Guilera, O.M.
\newblock Constraining mass, radius, and tidal deformability of compact stars
with axial wI modes: New universal relations including slow stable hybrid
stars.
\newblock {\em Mon. Not. R. Astron. Soc.} {\bf 2023}, {\em 519},~3194--3200. [\href{http://dx.doi.org/10.1093/mnras/stac3780}{CrossRef}]

\bibitem[Li et~al.(2025)Li, Watts, Zhang, Guillot, Xu, Santangelo, Zane, Feng,
Zhang, Ge, et~al.]{li:2025dmi}
Li, A.; Watts, A.L.; Zhang, G.; Guillot, S.; Xu, Y.; Santangelo, A.; Zane, S.;
Feng, H.; Zhang, S.N.; Ge, M.;  et~al.
\newblock Dense matter in neutron stars with eXTP.
\newblock {\em Sci. China Phys. Mech. Astron.} {\bf 2025}, {\em 68},~119503.
\bibitem[{Majczyna} et~al.(2020){Majczyna}, {Madej}, {Nale{\.z}yty},
{R{\'o}{\.z}a{\'n}ska}, and {Be{\l}dycki}]{Majczyna:2020pom}
{Majczyna}, A.; {Madej}, J.; {Nale{\.z}yty}, M.; {R{\'o}{\.z}a{\'n}ska}, A.;
{Be{\l}dycki}, B.
\newblock {Precision of Mass and Radius Determination for Neutron Stars from
the ATHENA Mission}.
\newblock {\em Astrophys. J. Lett.} {\bf 2020}, {\em 888},~123. [\href{http://dx.doi.org/10.3847/1538-4357/ab5dc9}{CrossRef}]
\end{thebibliography}
\end{document}